\documentclass[prd,superscriptaddress,altaffilletter,showpacs,nofootinbib]{revtex4}
\usepackage[dvips]{graphicx}
\usepackage{amsmath}
\usepackage{graphicx,epsfig}
\newcommand{\be}{\begin{equation}}
\newcommand{\ee}{\end{equation}}
\newcommand{\bea}{\begin{eqnarray}}
\newcommand{\eea}{\end{eqnarray}}
\newcommand{\der}{\partial}
\newcommand{\vphi}{\varphi}

\begin{document}


\title{Inflationary equilibrium configurations of scalar-tensor theories of gravity}


\author{Israel Quiros}\email{iquiros@fisica.ugto.mx}\affiliation{Dpto. Ingenier\'ia Civil, Divisi\'on de Ingenier\'ia, Universidad de Guanajuato, Gto., M\'exico.}

\author{Tame Gonzalez}\email{tamegc72@gmail.com}\affiliation{Dpto. Ingenier\'ia Civil, Divisi\'on de Ingenier\'ia, Universidad de Guanajuato, Gto., M\'exico.}

\author{Roberto De Arcia}\email{robertodearcia@gmail.com}\affiliation{Dpto. Astronom\'ia, Divisi\'on de Ciencias Exactas, Universidad de Guanajuato, Gto., M\'exico.}

\author{Ricardo Garc\'ia-Salcedo}\email{rigarcias@ipn.mx}\affiliation{CICATA-Legaria, Instituto Polit\'ecnico Nacional, Ciudad de México, CP 11500, México.}

\author{Ulises Nucamendi}\email{unucamendi@gmail.com}\affiliation{Instituto de F\'isica y Matem\'aticas, Universidad Michoacana de San Nicol\'as de Hidalgo, Edificio C-3, Ciudad Universitaria, CP. 58040 Morelia, Michoac\'an, M\'exico.}\affiliation{Mesoamerican Centre for Theoretical Physics, Universidad Aut\'onoma de Chiapas. Ciudad Universitaria, Carretera Zapata Km. 4, Real del Bosque (Ter\'an), 29040, Tuxtla Guti\'errez, Chiapas, M\'exico.}\affiliation{Departamento de F\'isica, Cinvestav, Avenida Instituto Polit\'ecnico Nacional 2508, San Pedro Zacatenco, 07360, Gustavo A. Madero, Ciudad de M\'exico, M\'exico.}

\author{Joel F. Saavedra}\email{joel.saavedra@ucv.cl}\affiliation{Instituto de F\'isica, Pontificia Universidad Cat\'olica de Valpara\'iso, Casilla 4950, Valpara\'iso, Chile.}

\date{\today}


\begin{abstract} In this paper we investigate the asymptotic dynamics of inflationary cosmological models that are based in scalar-tensor theories of gravity. Our main aim is to explore the global structure of the phase space in the framework of single-field inflation models. For this purpose we make emphasis in the adequate choice of the variables of the phase space. Our results indicate that, although single-field inflation is generic in the sense that the corresponding critical point in the phase space exists for a wide class of potentials, along given phase space orbits -- representing potential cosmic histories -- the occurrence of the inflationary stage is rather dependent on the initial conditions. We have been able to give quantitative estimates of the relative probability (RP) for initial conditions leading to slow-roll inflation. For the non-minimal coupling model with the $\phi^2$-potential our rough estimates yield to an almost vanishing relative probability: $10^{-13}\,\%\lesssim RP\ll 10^{-8}\,\%$. These bonds are greatly improved in the scalar-tensor models, including the Brans-Dicke theory, where the relative probability $1\,\%\lesssim RP\leq 100\,\%$. Hence slow-roll inflation is indeed a natural stage of the cosmic expansion in Brans-Dicke models of inflation. It is confirmed as well that the dynamics of vacuum Brans-Dicke theories with arbitrary potentials are non-chaotic.\end{abstract}


\maketitle


\section{Introduction}\label{sec-intro}

Due to its relative simplicity scalar fields represent a fruitful arena where to test several of the most relevant physical theories \cite{copeland-rev, bamba-rev}. These can be found in the low-energy limit of string theory which is equivalent to Brans-Dicke (BD) theory \cite{wands-rev, bd-1961} with the dilaton playing the role of the BD scalar field, as well as in models designed to explain the primordial inflation \cite{infl-linde-1, infl-linde-2, infl-linde-21, infl-linde-3, infl-linde-31, infl-guth, infl-albrecht, infl-reheat, accetta_ind_grav_infl} where the inflationary stage is driven by a self-interacting scalar field called as inflaton. Scalar fields may appear also as mediators of fundamental interactions as in the scalar-tensor theories (STT) of gravity \cite{fujii-book, faraoni-book, quiros-rev} where the gravitational interactions are mediated both by the metric and by a scalar field.

Of particular importance to understand several issues of standard cosmology such as flatness, horizon and monopole problems, is the inflationary paradigm \cite{infl-linde-1, infl-linde-2, infl-linde-21, infl-linde-3, infl-linde-31, infl-guth, infl-albrecht, infl-reheat, accetta_ind_grav_infl, infl-obs, infl-lyth, infl-n-gauss, liddle-prd-1994, barrow-prd-1995, lidsey-rmp-1995}. Single field inflation models are based on the possibility that a given scalar field $\phi$ -- the inflaton -- slowly rolls down its self-interaction potential $V(\phi)$. If the slow roll occurs (starts to be precise) in a regime of very high potential energy ($V\gg 1$), the resulting scenario is called as chaotic inflation \cite{lidsey-rmp-1995}. Multi-field inflation models represent an interesting alternative where two or several scalar fields may interact to produce the required amount of inflation \cite{assist-infl}. Hybrid inflation is, perhaps, the most successful model of the latter kind \cite{hyb-infl-linde, clesse-prd-2009}.

No matter whether one deals with multi-field or single field inflation models, their common feature is the high complexity of the mathematical structure: Only through the slow-roll approximation one may retrieve some useful analytic information on the inflationary dynamics. Otherwise one has to rely either on the numeric investigation or on the application of the tools of the dynamical systems theory \cite{zeldovich-1985, urena-ijmpd-2009, urena-prd-2016}. The use of the dynamical systems is specially useful when one deals with scalar-field cosmological models \cite{ellis-book, coley-book, wands-prd-1998, faraoni-grg-2013, bohmer-rev, quiros_ejp_rev, genly-1, genly-2, genly-3}. By means of the dynamical systems tools one may obtain very useful information on the asymptotic dynamics of the mentioned cosmological models. The asymptotic dynamics is characterized by: i) attractor solutions to which the system evolves for a wide range of initial conditions, ii) saddle equilibrium configurations that attract the phase space orbits in one direction but repel them in another direction, iii) source critical points which may be pictured as past attractors, or iv) limit cicles, among others.

At this point we have to mention that there is some tension between the existence of true attractor behavior and the Liouville's theorem \cite{carroll-prd-2013}. This result is correct only for Hamiltonian systems where the state space or phase space is spanned by 'canonically' conjugated variables.\footnote{We use the primes to mean that these are not really canonical variables.} Take, for instance, a self-interacting scalar field that is minimally coupled to gravity. The model is given by the action (throughout the paper we use the units where $8\pi G=M_\text{Pl}^{-2}=c=1$), $$S_\text{MC}=\int d^4x\sqrt{|g|}\left[\frac{1}{2}\,R-\frac{1}{2}(\der\phi)^2-V(\phi)\right],$$ where $R$ is the curvature scalar, $(\der\phi)^2\equiv g^{\mu\nu}\der_\mu\phi\der_\nu\phi$ and $V=V(\phi)$ is the self-interaction potential. In terms of homogeneous and isotropic Friedmann-Robertson-Walker (FRW) metric parametrized in the following way (we consider the case with flat spatial sections): $ds^2=-dt^2+e^{2\alpha(t)}\delta_{ij} dx^idx^j$, where $\alpha(t)$ is the time-dependent scale factor, the above action can be written as: $S_\text{MC}=\int d^3xdt{\cal L}_\text{MC}(\alpha,\dot\alpha,\phi,\dot\phi),$ where the Lagrangian density, $${\cal L}_\text{MC}=e^{3\alpha}\left[-3\dot\alpha^2+\frac{\dot\phi^2}{2}-V(\phi)\right],$$ and the overdot denotes derivative with respect to the cosmic time $t$. In this case one may introduce the following 'canonically' conjugated momenta: $$\pi_\alpha=\frac{\der{\cal L}_\text{MC}}{\der\dot\alpha}=-6e^{3\alpha}\dot\alpha,\;\pi_\phi=\frac{\der{\cal L}_\text{MC}}{\der\dot\phi}=e^{3\alpha}\dot\phi,$$ and to write the 'Hamiltonian' for this system: ${\cal H}_\text{MC}=e^{-3\alpha}\left[-\pi^2_\alpha+6\pi^2_\phi+12e^{6\alpha}V\right]/12.$ The equations of motion are then written in the form of Hamilton's equations:

\bea &&\dot\pi_\alpha=-\frac{\der{\cal H}_\text{MC}}{\der\alpha},\;\dot\alpha=\frac{\der{\cal H}_\text{MC}}{\der\pi_\alpha},\nonumber\\
&&\dot\pi_\phi=-\frac{\der{\cal H}_\text{MC}}{\der\phi},\;\dot\phi=\frac{\der{\cal H}_\text{MC}}{\der\pi_\phi},\nonumber\eea or, written in detail:

\bea &&\dot\alpha=-\frac{1}{6}e^{-3\alpha}\pi_\alpha,\;\dot\pi_\alpha=\frac{1}{4}e^{-3\alpha}\left[-\pi^2_\alpha+6\pi^2_\phi\right]-3e^{3\alpha}V,\nonumber\\
&&\dot\phi=e^{-3\alpha}\pi_\phi,\;\dot\pi_\phi=-e^{3\alpha}V_\phi,\label{hamilton-odes}\eea where $V_\phi\equiv\der V/\der\phi$. These four ordinary differential equations (ODE) together with the Hamiltonian constrain ${\cal H}_\text{MC}=0$, amount to the well known cosmological equations:

\bea 3\dot\alpha^2=\frac{\dot\phi^2}{2}+V,\;\;\ddot\alpha=-\frac{\dot\phi^2}{2},\;\;\ddot\phi+3\dot\alpha\dot\phi=-V_\phi.\label{cosmo-eq}\eea 

The phase space variables $\alpha$, $\phi$, $\pi_\alpha$ and $\pi_\phi$ are appealing because these allow to apply the Hamiltonian formalism to scalar-field cosmological models, so that one may endow the mathematical equations of the model with well-known physical meaning. However, scalar-field cosmological models admit in general more than one plausible set of phase space coordinates. In the above example, for instance, we may as well introduce the following dimensionless variables:\footnote{Notice that due to our choice of units where $8\pi G=M_\text{Pl}^{-2}=c=1$, the scalar field is a dimensionless quantity.}

\bea x=\frac{\dot\phi}{\sqrt{6}\,\dot\alpha},\;y=\frac{V_\phi}{V},\label{new-vars}\eea which are the phase space variables of certain two-dimensional (2D) state space. In terms of these new variables the cosmological equations \eqref{cosmo-eq} are traded by the following system of two ODE-s:

\bea \frac{dx}{d\alpha}=-\left(3x+\sqrt\frac{3}{2}\,y\right)(1-x^2),\;\;\frac{dy}{d\alpha}=\sqrt{6}\,xy^2\left(\Gamma-1\right),\label{2d-ode}\eea where now the scale factor $\alpha$ plays the role of 'time-ordering' variable or, simply, the time in the new state space and $\Gamma\equiv VV_{\phi\phi}/V^2_\phi$. For a wide class of self-interaction potentials, depending of the concrete functional form of the potential $V=V(\phi)$, the latter quantity is either a constant or it can be written as a function of the phase space variable $y$: $\Gamma=\Gamma(y)$. It can be shown, in particular, that $\Gamma=1$ for the exponential potential $V\propto\exp{(\sigma\phi)}$, while for the quadratic potential ($V\propto\phi^2$) $\Gamma=1/2$ (for the quartic potential $V\propto\phi^4$, $\Gamma=3/4$). For other potentials it can be a function instead. For the symmetry-breaking potential $V=\lambda(\phi^2-\mu^2)^2/4$, for instance: $$\Gamma=\frac{4\left[3\phi^2_\pm(y)-\mu^2\right]}{\phi^2_\pm(y)-\mu^2},\;\phi_\pm(y)=\frac{2}{y}\left(1\pm\sqrt{1+\frac{\mu^2y^2}{4}}\right).$$ In \eqref{2d-ode} we used the Friedmann constraint (first equation in \eqref{cosmo-eq}): 

\bea \frac{V}{3\dot\alpha^2}=1-x^2,\label{hamilton-c}\eea in order to eliminate the term with the potential from the equations.

Then we have -- at least -- two different dynamical systems that amount to different plausible phase space representations of the cosmological equations \eqref{cosmo-eq}: i) a three-dimensional (3D) Hamiltonian dynamical system on the variables $\alpha$, $\phi$, $\pi_\phi$,

\bea \dot\alpha=\mp\frac{e^{-3\alpha}}{6}\sqrt{6\pi^2_\phi+12e^{6\alpha}V},\;\;\dot\phi=e^{-3\alpha}\pi_\phi,\;\;\dot\pi_\phi=-e^{3\alpha}V_\phi,\label{3d-ode}\eea where the Hamiltonian constraint: $\pi_\alpha=\pm\sqrt{6\pi^2_\phi+12e^{6\alpha}V},$ has been explicitly considered in order to eliminate the variable $\pi_\alpha$, and ii) the 2D non-Hamiltonian dynamical system \eqref{2d-ode} on the variables $x$, $y$. In addition to the different dimensionality, the structure of the phase space is different in both cases so that these are not equivalent. Actually, the only critical point of \eqref{3d-ode}, $$\dot\alpha=\dot\phi=\dot\pi_\phi=0\;\Rightarrow\;V=0,\;V_\phi=0,$$ is the attractor point corresponding to a static universe ($\alpha=\alpha_0=$const.) with the scalar field sitting on a extremum of $V$ where the potential vanishes. I. e., in this equilibrium configuration the scalar field has vanishing energy since both its kinetic energy $\propto\dot\phi^2$, and its potential energy density $V$, vanish. Meanwhile, in the general case without specifying the functional form of the potential $V$, the dynamical system \eqref{2d-ode} has three critical points, $P_i:(x_i,y_i)$, that are associated with non-trivial cosmological evolution,\footnote{For both choices of phase space variables the mentioned equilibrium points do not exhaust the whole phase space structure since there can be critical points that are located at infinities. However, for the purposes of the present discussion it is not necessary to take into account the whole phase space structure.} $x'=y'=0$: 1) the stiff-mater solutions, $$P_\text{stiff}:(\pm 1,0)\;\Rightarrow\;3\dot\alpha^2=\frac{1}{2}\,\dot\phi^2\;\Rightarrow\;\alpha(t)=\pm\frac{\phi(t)}{\sqrt{6}}+C_0,$$ where $C_0$ is an arbitrary integration constant, and 2) de Sitter expansion, $$P_\text{dS}:(0,0)\;\Rightarrow\;\dot\alpha=\sqrt\frac{V_0}{3}\;\Rightarrow\;\alpha(t)=\sqrt\frac{V_0}{3}\,t+C_0,$$ where $C_0$ and $V_0$ are constants. This means that there is a clear difference between the above choices of the phase space variables. The dimensionality of the phase space, which is different for the above choices, plays a fundamental role \cite{fang-2016} in particular in the search for chaotic behavior. Recall that, based on the Poincar\`e-Bendixson theorem \cite{p-b-theor-book-1, p-b-theor-book-2, p-b-theor-book-3, p-b-theor-book-4}, it can be concluded that chaos may arise only in phase spaces with dimension higher than two. Hence, an adequate choice of the phase space variables is of particular importance in the discussion about possible chaotic behavior in scalar-field cosmological models \cite{calzetta-cqg-1993, bombelli-ijmp-1998, giacomini-prd-2001, faraoni-cqg-2006}.

Given that the choice of variables in \eqref{3d-ode} leads to trivial static space solution and that, besides, due to Liouville's theorem the true attractor character of the solution is unclear \cite{carroll-prd-2013}, in the present paper, as in most papers on the study of scalar-field cosmological models \cite{copeland-rev, ellis-book, coley-book, wands-prd-1998, faraoni-grg-2013, bohmer-rev, quiros_ejp_rev, genly-1, genly-2, genly-3}, we choose variables of the phase space that do not lead to Hamiltonian dynamics, so that we do not have to care about Liouville's theorem. What one really should care about are those variables that: i) are dimensionless so that the results of the analysis do not depend on the chosen units, ii) span the phase space with the lowest dimensionality, iii) are able to cover the whole phase space and iv) are bounded. The latter requirements are very important when one aims at establishing a quantitative measure for determining the relative probability of initial conditions leading to a given critical point (or to its neighborhood). Actually, if the given variables cover the whole phase space and are bounded, then the phase space is finite and one may compute its volume and/or of any of its subsets, sot that a geometric probability can be established. 

The aim of the present paper is twofold. On one hand, there is a widespread belief that on the basis of the dynamical systems analysis it may be concluded that inflation is a fairly general property of solutions of scalar-field models \cite{zeldovich-1985, urena-ijmpd-2009}. Such a conclusion may be correct if the inflationary behavior can be associated with attractor critical points in the phase space so that, no matter what initial conditions to choose, the corresponding phase space orbits, representing plausible cosmological dynamics, are attracted towards the inflationary point. However, due to the unclear identification of those critical points that may be associated with inflationary behavior in the bibliography (see, for instance, the discussion in \cite{urena-ijmpd-2009}), it is interesting to revise the mentioned result. On the other hand, despite of the widespread use of the dynamical systems tools in cosmology \cite{copeland-rev, zeldovich-1985, urena-ijmpd-2009, urena-prd-2016, ellis-book, coley-book, wands-prd-1998, faraoni-grg-2013, bohmer-rev, quiros_ejp_rev, genly-1, genly-2, genly-3}, there are few published studies where the inflationary dynamics are correlated with equilibrium points in some phase space, and where a quantitative measure of the amount of initial conditions leading to the inflationary critical points is discussed. In this regard, here we want to explore the global phase space dynamics of single-field scalar-tensor models of inflation, with emphasis in the search for critical points that may be correlated with the inflationary dynamics. Then, on the basis of geometric probability, we shall establish a quantitative measure of the amount of initial conditions leading to the inflationary equilibrium points. This will allow us to give quantitative estimates of the relative probability of slow-roll inflation in the models. 

Here we shall focus in scalar-tensor theories of gravity that are based in the following action:\footnote{Notice that under the innocuous scalar field redefinition $\phi\rightarrow\vphi:\phi=h(\vphi)$, the gravitational piece of the action \eqref{action} can be recast into the alternative form: $$S=\frac{1}{2}\int d^4x\sqrt{-g}\left[F(\vphi)R-\omega(\vphi)(\der\vphi)^2-2V(\vphi)\right],$$ where $\omega(\vphi)\equiv h^2(\vphi)$.}

\bea S=\frac{1}{2}\int d^4x\sqrt{-g}\left[FR-(\der\phi)^2-2V+2{\cal L}_m\right],\label{action}\eea where $F=F(\phi)$ is an arbitrary non-negative function of the scalar field, $V=V(\phi)$ is its self-interacting potential and ${\cal L}_m$ is the Lagrangian of the matter degrees of freedom other than the scalar field (radiation, baryons, cold dark matter, etc). The equations of motion resulting from \eqref{action} read:

\bea &&FG_{\mu\nu}=T^{(m)}_{\mu\nu}+\der_\mu\phi\der_\nu\phi-\frac{1}{2}g_{\mu\nu}(\der\phi)^2-g_{\mu\nu}V+F_{\phi\phi}\left[\der_\mu\phi\der_\nu\phi-g_{\mu\nu}(\der\phi)^2\right]+F_\phi\left(\nabla_\mu\nabla_\nu\phi-g_{\mu\nu}\nabla^2\phi\right),\nonumber\\
&&\nabla^2\phi+\frac{F_\phi(1+3F_{\phi\phi})}{2F+3F^2_\phi}\,(\der\phi)^2=\frac{F_\phi T_{(m)}}{2F+3F^2_\phi}+\frac{2FV_\phi-4F_\phi V}{2F+3F^2_\phi},\label{action-mot-eq}\eea where $T^{(m)}_{\mu\nu}=-2\delta(\sqrt{-g}{\cal L}_m)/\delta g^{\mu\nu}$ is the stress-energy tensor of matter, $T_{(m)}=g^{\mu\nu}T^{(m)}_{\mu\nu}$ its trace and $\nabla^2\equiv g^{\mu\nu}\nabla_\mu\nabla_\nu$. Besides, we are using the following notation: $X_\phi\equiv\partial X/\partial\phi$, $X_{\phi\phi}\equiv\partial^2X/\partial\phi^2$, etc.

We have organized the paper in the following way. In section \ref{sect-phi2-infl} we make a detailed revision of the dynamical systems investigation of FRW cosmological models of a massive scalar field. No matter how simple the model seems, the existing papers did not succeed in identifying any actual critical points that may be associated with inflationary dynamics. In this section, through using appropriate variables of the phase space, we identify the critical points that are correlated with slow-roll inflation. In section \ref{sect-stt} we discuss on the inflationary dynamics of FRW cosmological models that are based in scalar-tensor theories of gravity of the type \eqref{action}. We obtain generic critical points that are quite independent of the coupling function $F(\phi)$ and of the potential $V(\phi)$. Among these we identify points that belong in a critical manifold that corresponds to the slow-roll inflation. Then we focus in particular coupling functions, as the one for non-minimal coupling (NMC) theories: $F=1-\epsilon\phi^2$, in section \ref{sect-nmc} and for Brans-Dicke theory: $F=\epsilon\phi^2$, in section \ref{sect-bd}. The results of the present study are discussed in detail in section \ref{sect-discuss}. In this section we focus mainly in the question on how natural primordial inflation really is in what regards to the required initial conditions. For this purpose we define a rough quantitative measure that is based on geometric probability. We are able to do this thanks to the correct choice of phase space variables. The possibility of chaotic behavior of the scalar field dynamics in the phase space is also briefly discussed. In section \ref{sect-conclu} concluding remarks are given.


\section{Dynamical systems study of $\phi^2$-inflation}\label{sect-phi2-infl}

In this section we shall revise the dynamical systems study of FRW cosmological models with a minimally coupled massive scalar field also known as $\phi^2$-inflation \cite{zeldovich-1985, urena-ijmpd-2009}. The action of the model is given by:

\bea S=\frac{1}{2}\int d^4x\sqrt{|g|}\left[R-(\der\phi)^2-m^2\phi^2\right],\label{phi2-action}\eea where $m$ is the mass of the scalar (inflaton) field. In an FRW spacetime with flat spatial sections, whose line element is given by:\footnote{Here and for the rest of the paper we use a different parametrization for the metric than the one in the introductory part.} 

\bea ds^2=-dt^2+a^2(t)\delta_{ik}dx^idx^k,\label{frw-metric}\eea the equations of motion read:

\bea &&H^2=\frac{1}{6}\left(\dot\phi^2+m^2\phi^2\right),\nonumber\\
&&\dot H=-\frac{1}{2}\dot\phi^2,\nonumber\\
&&\ddot\phi=-3H\dot\phi-m^2\phi.\label{feq}\eea The slow-roll conditions that are necessary for early inflation to occur, amount to \cite{liddle-prd-1994, barrow-prd-1995, lidsey-rmp-1995}:

\bea |\ddot\phi|\ll H|\dot\phi|,\;\;\dot\phi^2\ll V.\label{slow-roll-c}\eea 

From the dynamical systems perspective this model have been studied for the first time in \cite{zeldovich-1985} and since then similar qualitative studies have been performed \cite{urena-ijmpd-2009}. In the pioneering work \cite{zeldovich-1985} the authors chose a mix of dimensionful and dimensionless variables: $\bar x=\phi/3m$, $\bar y=\dot\phi/3$ and $\bar z=H/m$. Their analysis did not reveal inflationary attractor solutions but rather asymptotic inflationary behavior at $t\rightarrow-\infty$ that is associated with unstable critical points. In \cite{urena-ijmpd-2009} the following set of dimensionless variables, $$\hat x=\frac{\dot\phi}{\sqrt{6}\,H},\;\hat y=\frac{m\phi}{\sqrt{6}\,H},\;\hat z=\frac{m}{H},$$ was used. The authors of this work took the projection of the phase space into the $\hat x\hat y$-plane as the meaningful space where to look for inflationary behavior and regarded the variable $\hat z$ as a 'control parameter'. In that paper, as in the former \cite{zeldovich-1985}, it was not possible to find actual critical points that could be associated with the inflationary dynamics.\footnote{The authors of REF. \cite{urena-ijmpd-2009} themselves recognize that, what they called as scalar-field dominated critical points, were not fixed points in the strict sense since these depended on the third phase space coordinate $\hat z$.} Despite of the lack of success in the identification of equilibrium states in the phase space that could be associated with inflationary dynamics, the authors of \cite{zeldovich-1985} came to the conclusion that the inflationary stage is a fairly general property of the massive scalar field model.\footnote{There are divided opinions on whether or not single-field inflation arises naturally \cite{infl-linde-21, inic-infl-1, inic-infl-2, inic-infl-3, inic-infl-4, inic-infl-5}.} Here we want to revise this result by correctly identifying critical points that represent slow-rolling inflationary behavior.


The first step is to identify appropriate variables of some state space. If introduce the following dimensionless and bounded variables:

\bea x=\frac{\dot\phi}{\sqrt{6}H},\;y=\frac{m\phi}{\sqrt{6}H},\;z=\frac{H}{H+m},\label{ps-vars}\eea where $|x|\leq 1$, $|y|\leq 1$ and $0\leq z\leq 1$ (we consider only expanding cosmologies so that $H\geq 0$). Due to the Friedmann constraint 

\bea x^2+y^2=1\;\Rightarrow\;y=\pm\sqrt{1-x^2},\label{fried-c-0}\eea one of the above variables is not independent from the others. 

The two-dimensional dynamical system corresponding to the cosmological equations \eqref{feq} is given by the following pair of ordinary differential equations on the independent variables $x$, $z$:

\bea &&x'_\pm=\sqrt{1-x^2}\left[-3x\sqrt{1-x^2}z\mp(1-z)\right],\nonumber\\
&&z'=-3x^2z^2(1-z),\label{2d-dyn-syst}\eea where the comma denotes derivative with respect to the time variable $\tau=mt+\ln a$, and the '$\pm$' signs account for two branches of the dynamical system. Our variables $x$, $z$ are bounded so that the global asymptotic dynamics of the $\phi^2$-inflation model is contained within the rectangle: $\Psi=\{(x,z)|-1\leq x\leq 1, 0\leq z\leq 1\}$. This means that we do not have to apply the procedure based on the projection of points at infinity onto the equator of the Poincar\`e sphere, as in \cite{zeldovich-1985} (see also \cite{bohmer-rev, genly-plb-2007}).

It will be useful to write the slow-roll conditions \eqref{slow-roll-c} in terms of the phase space variables \eqref{ps-vars}:

\bea |\ddot\phi|\ll H|\dot\phi|\rightarrow 3H\dot\phi\simeq-m^2\phi\rightarrow z_\pm\simeq\frac{\sqrt{1-x^2}}{\sqrt{1-x^2}\mp 3x},\label{xyz-slow-roll-c1}\eea and

\bea \dot\phi^2\ll V\rightarrow 3H^2\simeq\frac{1}{2}\,m^2\phi^2\rightarrow y^2\simeq 1\rightarrow x\simeq 0.\label{xyz-slow-roll-c2}\eea The first of the slow-roll conditions above is represented by an epsilon-neighborhood around the curves $z_\pm=z_\pm(x)$, while the second one is depicted by the epsilon-neighborhood of the point $x=0$. Notice that both slow-roll conditions coincide in the epsilon-neighborhood of the phase space point $(x,z)=(0,1)$.


\begin{figure}
\includegraphics[width=7cm, height=7.5cm]{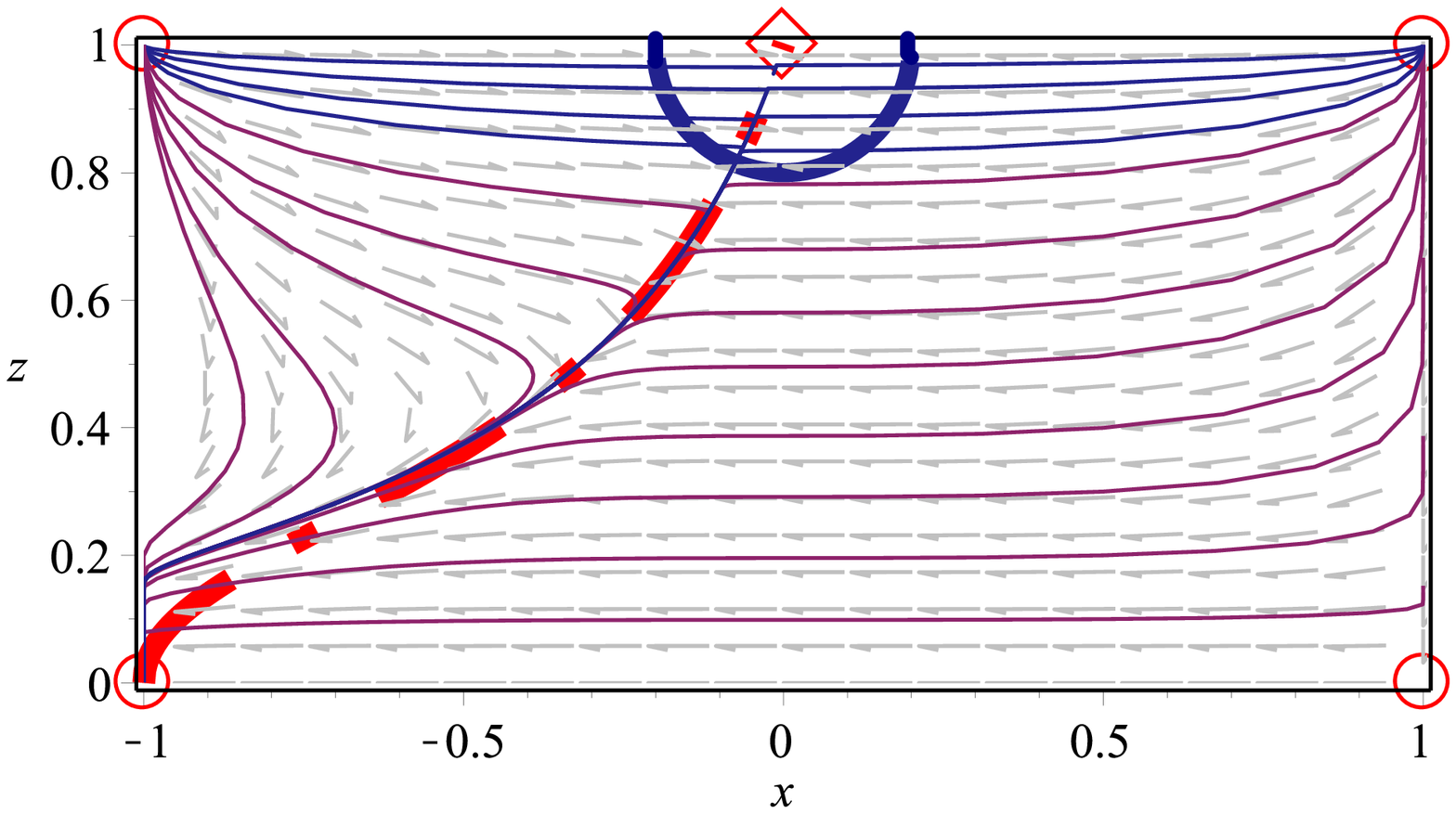}
\includegraphics[width=7cm, height=7.5cm]{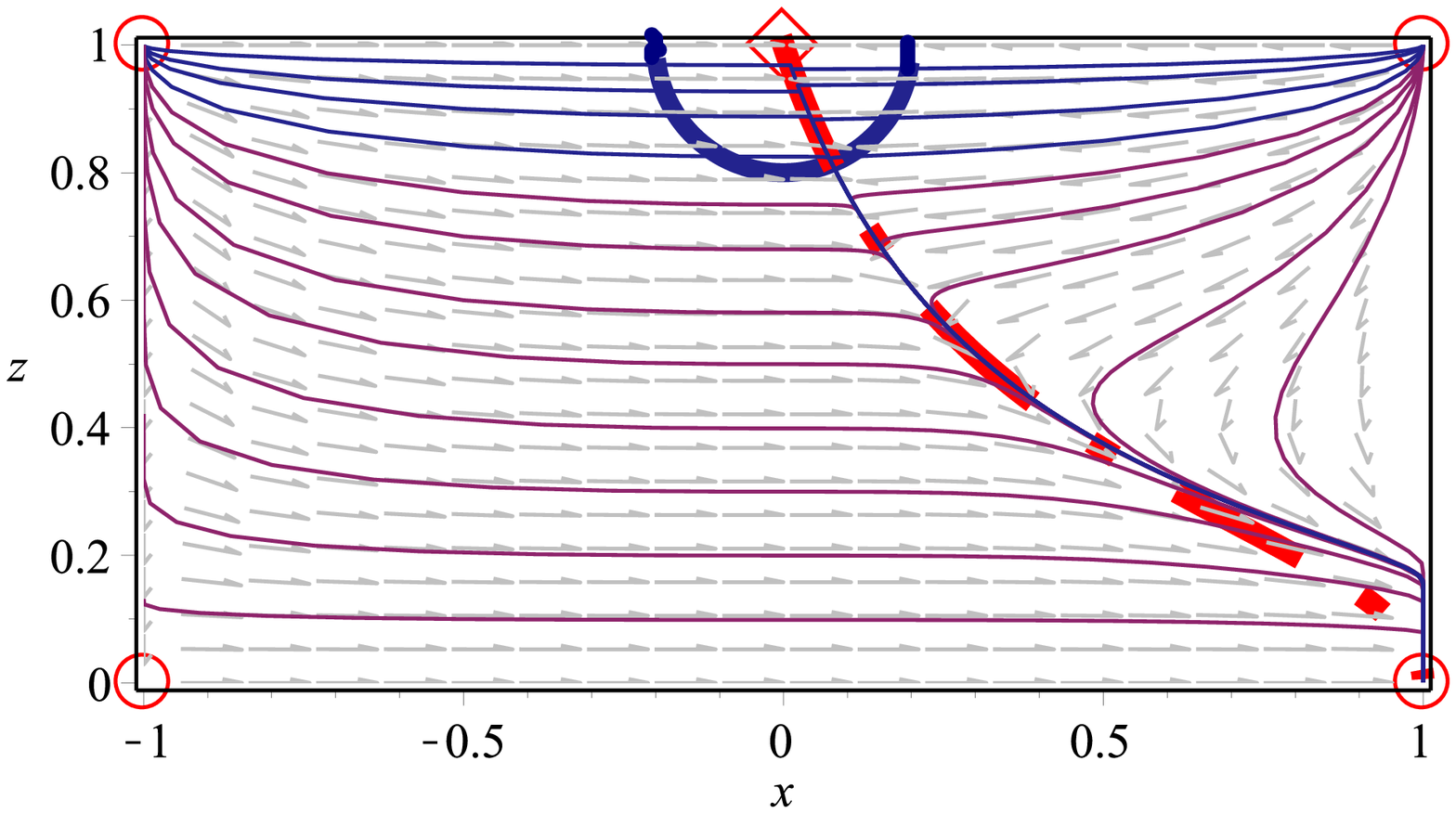}
\vspace{1cm}\caption{Phase portrait of the dynamical system \eqref{2d-dyn-syst} for the positive and negative branches (left and right hand panels respectively). The small circles in the corners of each phase rectangle enclose the critical points of the dynamical system, but for the diamond that encloses the saddle critical point $P_\text{inf}:(0,1)$, which is the one associated with primordial inflation. The thick dash-dot curves correspond to the slow-roll condition \eqref{xyz-slow-roll-c1}: $z_\pm=\sqrt{1-x^2}/(\sqrt{1-x^2}\pm 3x)$. These joint the inflationary saddle point $P_\text{inf}$ with the future attractor: the stiff matter solution (points $(-1,0)$ in the left hand panel and $(1,0)$ in the right hand panel). The slow-roll conditions \eqref{xyz-slow-roll-c1} and \eqref{xyz-slow-roll-c2} are jointly satisfied in the neighborhood of the inflationary point. The thick solid circle that encloses the diamond represents a $\rho$-ball around the slow-roll inflation point. The ball of radius $\rho$ is hit by those orbits that stay enough time in the neighborhood of the de Sitter point as to produce the required amount of inflation.}\label{fig1}\end{figure}



The critical points $P_i:(x_i,z_i)$ of the positive branch of the dynamical system \eqref{2d-dyn-syst} -- phase portrait drawn in the left hand panel of FIG. \ref{fig1} -- are:

\begin{itemize}

\item Four stiff-matter equilibrium points: the past attractors $(\pm 1,1)$, the saddle point $(1,0)$ and the future attractor $(-1,0)$. The latter points are associated with a static universe ($H=0$), while the former ones correspond to $H\gg m$. All of these critical points correspond to stiff matter solutions since $x=\pm 1\Rightarrow 3H^2=\dot\phi^2/2$.

\item The saddle point corresponding to primordial de Sitter inflation, $P_\text{dS}:(0,1)$. In this case $x\rightarrow 0\Rightarrow\phi\approx\phi_0=$const., i. e., $$H\approx H_0=\frac{m\phi_0}{\sqrt{6}}.$$ The condition $z=1$ implies either the formal limit $m\rightarrow 0$ -- massless scalar field -- or $H\gg m$ which, in turn, may be associated with large values $\phi_0\gg 1$. Therefore, the de Sitter equilibrium configuration $P_\text{dS}$ is attained at large field values, far from the minimum of the potential.

\end{itemize} The negative branch of the dynamical system (phase portrait drawn in the right hand panel of FIG. \ref{fig1}) has the same equilibrium points. The only difference is in the stability properties of the points $(\pm 1,0)$. In this case $(-1,0)$ is a saddle while $(1,0)$ is the future attractor.

Unlike former works \cite{zeldovich-1985, urena-ijmpd-2009}, here we have been able to find the critical point $P_\text{dS}$ that is associated with inflationary behavior. This equilibrium state is easily recognizable since, as seen from FIG. \ref{fig1}, the slow-roll conditions \eqref{xyz-slow-roll-c2}: $$x\approx 0,\;z_\pm=\frac{\sqrt{1-x^2}}{\sqrt{1-x^2}\mp 3x}\approx 1,$$ are jointly met, precisely, at the inflationary de Sitter point $(0,1)$ -- enclosed within the small diamond in the figure. From the phase portrait it is seen, also, that all of the phase plane orbits emerge from either of the stiff-matter past attractors $(-1,1)$ or $(1,1)$. The fact that the inflationary equilibrium state is a saddle critical point instead of an attractor entails that: i) inflation in this model is not as generic as concluded, for instance, in REF. \cite{zeldovich-1985} and ii) inflation is a transient stage of the cosmic evolution so that exit from inflation is a natural phenomenon. The fact that the attractor solution is the stiff-matter point $H=-\dot\phi/\sqrt{6}$, with $H\approx 0$ means that the end-point of the expansion is a static universe. This is due of course to the fact that the present model is not designed to describe the late-time dynamics of our universe. In subsection \ref{subsect-lambda} we shall modify the model by including a cosmological constant term and we shall look for the corresponding modification of the late-time dynamics.


\subsection{Comparison of our study with similar studies}

It is seen that our variables \eqref{ps-vars} do not differ too much from the variables of \cite{urena-ijmpd-2009}: $x=\hat x$, $y=\hat y$, $z=1/(1+\hat z)$. Then, why the authors of that reference did not find the correct critical point that is associated with the inflationary behavior? The major difference of our study is in the correct identification of the independent variables that span the physically meaningful 2D phase space: the variables $x$, $z$, instead of $\hat x$, $\hat y$ that are actually dependent of each other due to the Friedmann constraint: $\hat x^2+\hat y^2=1$. In terms of the variables $\hat x$, $\hat y$, $\hat z$, the dynamical system corresponding to the cosmological equations \eqref{cosmo-eq} reads (equations (5a), (5b) and (5c) of \cite{urena-ijmpd-2009}):

\bea &&\hat x'=-3\hat x(1-\hat x^2)-\hat y\hat z,\nonumber\\
&&\hat y'=3\hat x^2\hat y+\hat x\hat z,\nonumber\\
&&\hat z'=3\hat x^2\hat z,\label{luis-3d-ode}\eea where the comma is for derivative with respect to the number of e-foldings $N=\ln a$. Notice that if in the first and second equations above we substitute the Friedmann constraint $\hat y=\pm\sqrt{1-\hat x^2}$, where $$\hat y'=-\frac{\hat x\hat x'}{\pm\sqrt{1-\hat x^2}},$$ we obtain one and the same equation, so that only one of them is an independent ordinary differential equation. Let us to choose the first one. Then we are left with the following 2D dynamical system:

\bea &&\hat x'=-3\hat x(1-\hat x^2)\pm\hat z\sqrt{1-\hat x^2},\nonumber\\
&&\hat z'=3\hat x^2\hat z,\label{luis-2d-ode}\eea where the '$\pm$' signs account for two possible branches as in \eqref{2d-dyn-syst}. This is the physically meaningful dynamical system and the phase space where to look for the critical points of \eqref{luis-2d-ode} is the semi-infinite plane $\hat\Psi=\{(\hat x,\hat z)|-1\leq\hat x\leq 1,0\leq\hat z<\infty\}$. In this regard it is not difficult to see that the choice of the phase plane $\Psi_*=\{(\hat x,\hat y)|-1\leq\hat x\leq 1,0\leq\hat y\leq 1,\hat x^2+\hat y^2=1\}$ in \cite{urena-ijmpd-2009} as the physically meaningful region where to look for critical points, is an illusion since, due to the Friedmann constraint $\Psi_*$ is not a phase plane but a unit circle on the plane $\hat x\hat y$. There are three critical points of the dynamical system \eqref{luis-2d-ode} in the phase plane $\hat\Psi$: i) the stiff-matter source points $(\hat x,\hat z)=(\pm 1,0)$ and ii) the inflationary saddle point at the origin $(0,0)$. Other two stiff-matter critical points are located at $\hat z\rightarrow\infty$, so that projection of points at infinity onto the equator of the Poincar\`e sphere is required in order to complement the investigation in the $\hat x\hat z$-plane. The critical point at the origin $(0,0)$ in the $\hat x$, $\hat z$- coordinates is equivalent to our $P_\text{dS}:(0,1)$ above in terms of $x$, $z$, so that it is in fact the inflationary critical point the authors of \cite{zeldovich-1985, urena-ijmpd-2009} were searching for.


\begin{figure}
\includegraphics[width=7cm]{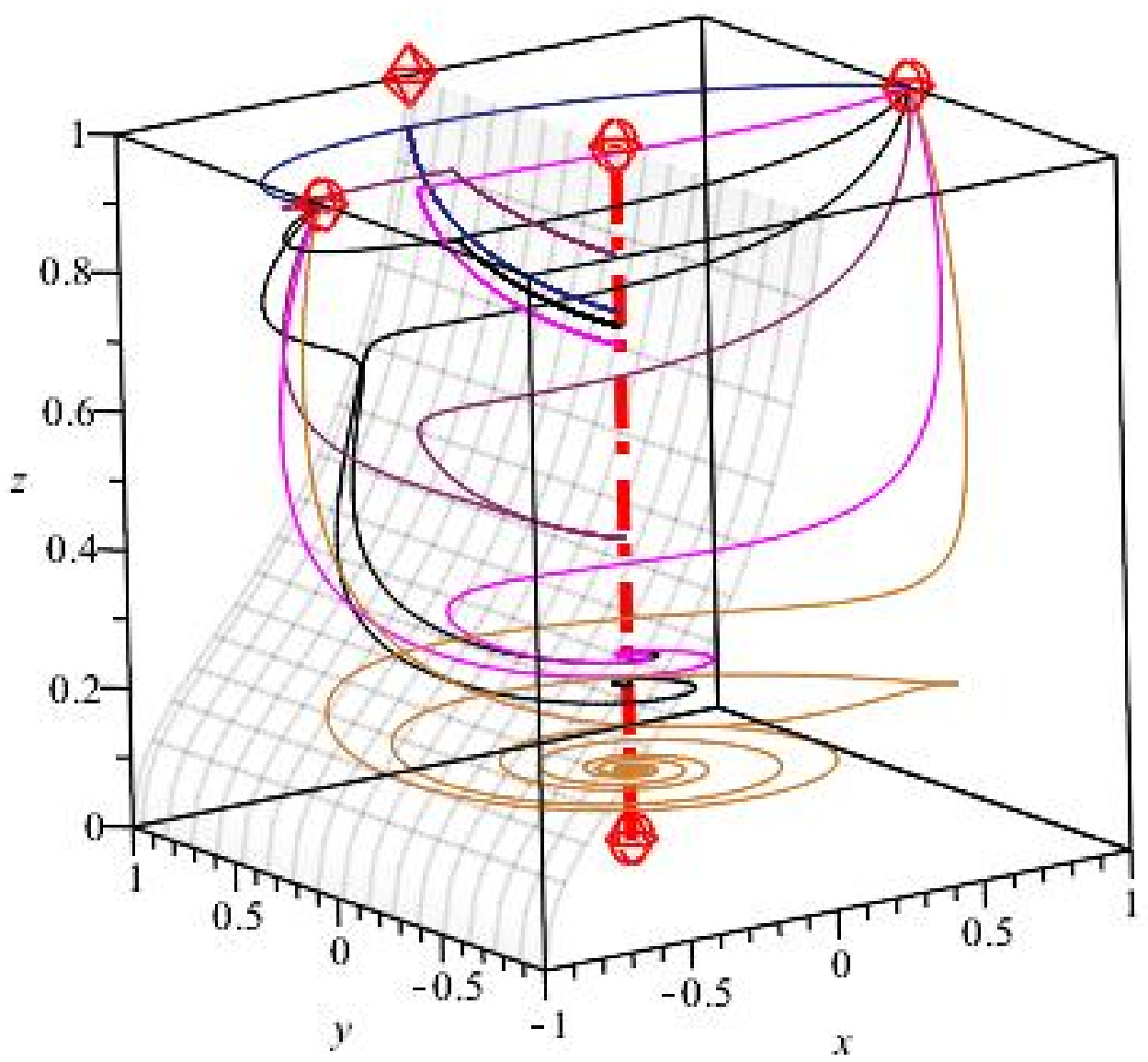}
\includegraphics[width=7cm]{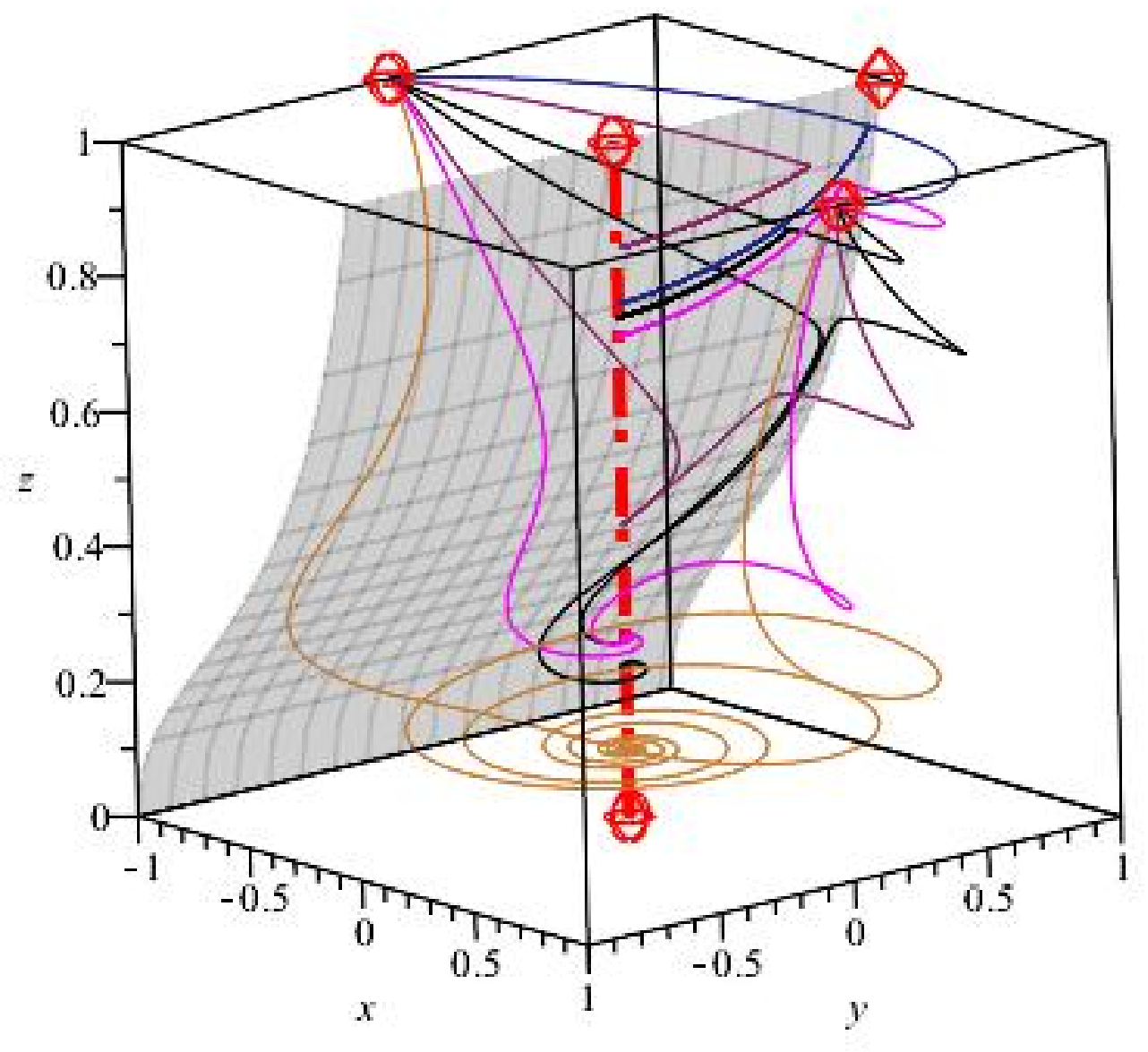}
\vspace{1cm}\caption{Phase portrait of the dynamical system \eqref{3d-dyn-syst}. The different views from the directions $(\theta,\vphi,\psi)=(-130,75,-5)$ and $(-45,75,-5)$ are shown. The small circles enclose the critical points of the dynamical system but for the inflationary saddle point $P_\text{inf}:(0,1,1)$, which is enclosed by the diamond. The dash-dot straight line ${\cal P}_\text{ds}:(0,0,z)$ represents the de Sitter attractor manifold. The gridded surface corresponds to the slow-roll condition \eqref{xyz-slow-roll-c1}: $z=\sqrt{1-x^2}/(\sqrt{1-x^2}-3x)$. The region in the phase portrait where the slow-roll conditions \eqref{xyz-slow-roll-c1} and \eqref{xyz-slow-roll-c2} are jointly satisfied corresponds to a small neighborhood of the inflationary point $P_\text{inf}$ lying on the gridded surface.}\label{fig2}\end{figure}



\subsection{$\phi^2$-inflation plus a cosmological constant}\label{subsect-lambda}

If in the action \eqref{phi2-action} add a Lagrangian piece corresponding to a cosmological constant term: $2\Lambda$, then the dynamical system corresponding to \eqref{feq} is the following 3D system of autonomous ODE-s:

\bea &&x'=-3x(1-x^2)z-y(1-z),\nonumber\\
&&y'=x(1+3xyz-z),\nonumber\\
&&z'=-3x^2z^2(1-z),\label{3d-dyn-syst}\eea where, as above, the comma denotes derivative with respect to the number of e-foldings $N=\ln a$, and the constraint $x^2+y^2=1$ has been replaced by the modified Friedmann constraint:

\bea \Omega_\Lambda=1-x^2-y^2,\label{modif-friedm-const}\eea with $\Omega_\Lambda=\Lambda/3H^2$ -- the dimensionless energy density of the cosmological constant ($\Omega_\Lambda\geq 0$). 

The physically meaningful phase space where to search for critical points of \eqref{3d-dyn-syst} is the cylinder: $$\Psi_\Lambda=\left\{(x,y,z)|-1\leq x\leq 1,-1\leq y\leq 1,x^2+y^2\leq 1,0\leq z\leq 1\right\}.$$ The critical points $P_i:(x_i,y_i,z_i)$ of \eqref{3d-dyn-syst} in $\Psi_\Lambda$ are listed below (see FIG. \ref{fig2}). 

\begin{itemize}

\item De Sitter attractor manifold ${\cal P}_\text{dS}:(0,0,z),$ for which $\Omega_\Lambda=1$, $q=-1$ and $$\lambda=\left\{0,-\frac{3}{2}z\pm\frac{1}{2}\sqrt{5z^2+8z-4}\right\}.$$ The vanishing eigenvalue of the linearization matrix is associated with the eigenvector that is tangent to the manifold ${\cal P}_\text{dS}$ at each point. The non-vanishing (conjugated) eigenvalues are always negative. Actually, $\lambda_-=-3z-\sqrt{5z^2+8z-4}$ is clearly negative, meanwhile, the negativity of $\lambda_+=-3z+\sqrt{5z^2+8z-4}$ is achieved whenever $(z-1)^2>0$, which is always satisfied.

\item Stiff-matter solutions $P^\pm_\text{sm}:(\pm 1,0,1)$, with $q=2$, $\lambda=\{6,3,3\}$, represent isolated source critical points (past attractors).

\item Slow-roll inflation solution $P_\text{inf}:(0,1,1)$ $\Rightarrow\Omega_\Lambda=0$, with $q=-1$, $\lambda=\{0,0,3\}$. Given that two of the eigenvalues of the linearization matrix are vanishing, nothing can be said about the stability of this critical point. The numeric computations show that this is in fact an isolated saddle point.

\end{itemize} Then, as expected, the addition of a cosmological term modified the late-time dynamics by replacing the stiff-matter (quasi-static) attractor in the classical vacuum case with the de Sitter attractor associated with the energy density of the quantum vacuum, i. e., the cosmological constant $\Lambda$. In this case two inflationary stages coexist together: the slow-roll (primordial) inflation stage associated with the saddle de Sitter point $P_\text{inf}$ and the late-time inflation states that belong in the attractor manifold ${\cal P}_\text{dS}$.


\subsection{Arbitrary potentials}

For arbitrary potentials the cosmological equations read:

\bea &&3H^2=\frac{\dot\phi^2}{2}+V(\phi),\nonumber\\
&&\dot H=-\frac{1}{2}\dot\phi^2,\nonumber\\
&&\ddot\phi=-3H\dot\phi-V_\phi.\label{feq-gen}\eea In this case it is convenient to introduce the following phase space variables:

\bea x=\frac{\dot\phi}{\sqrt{6}H},\;y=\frac{\sqrt{V}}{\sqrt{3}H},\;v=\frac{V_\phi}{V}.\label{ph-vars}\eea In terms of these variables the Friedmann constraint looks the same as above: $x^2+y^2=1$, so that only one of these variables is an independent phase space variable. We choose $x$. The cosmological equations \eqref{feq-gen}
 are traded by the following 2D dynamical system:

\bea &&x'=-\sqrt\frac{3}{2}\,(1-x^2)(\sqrt{6}\,x+v),\nonumber\\
&&v'=\sqrt{6}\,\alpha xv^2,\label{ode-gen}\eea where $\alpha=\Gamma_V-1$ with $\Gamma_V=VV_{\phi\phi}/V^2_\phi$. The slow-roll conditions \eqref{slow-roll-c} amount to:

\bea \left|\frac{\dot H}{H^2}\right|\ll 1\Rightarrow 3x^2\ll 1\Rightarrow x^2\approx 0,\;\frac{\ddot\phi}{H\dot\phi}\ll 1\Rightarrow v\approx-\frac{\sqrt{6}\,x}{1-x^2}\rightarrow 0.\label{slow-r-gen}\eea These conditions are jointly satisfied at the origin $(x,v)=(0,0)$, where $\phi\rightarrow\phi_0$ -- a constant, and $v\rightarrow 0$, i. e., $V_\phi/V\rightarrow 0$. The corresponding critical point $P_\text{inf}:(0,0)$ is the one to be associated with the slow-rolling inflationary dynamics: $H=H_0=\sqrt{V/3}$. The bond $v\approx 0$ restricts the kind of potentials that drive the inflationary stage. The exponential potential $V=V_0\exp(\sigma\phi)$, for instance, does not lead to slow-roll inflation since, in this case: $$v=\frac{V_\phi}{V}=\sigma,$$ so that the critical point $P_\text{inf}$ exists only if $\sigma\rightarrow 0$, i. e., for the constant potential or at large field values for the exponential potential with a very flat tail. 

In general the slow-roll inflationary critical point $P_\text{inf}$ exists for potentials with extrema such that the potential is non-vanishing (at least) at one of the extrema and/or at very large field values far from the extrema. Take, for instance, the symmetry-breaking type of potential $$V(\phi)=\frac{\lambda}{4}\left(\phi^2-\mu^2\right)^2,$$ where $\lambda$ is a dimensionless constant and $\mu$ is the mass parameter. In this case, $$v=\frac{4\phi}{\phi^2-\mu^2},$$ so that, $v\rightarrow 0$, implies either that $\phi=0$, where the potential is a local maximum, $V_\text{max}=V(0)=\lambda\mu^4/4$, or that the critical point is approached at very large field values, $\phi\gg\mu$, where $$V(\phi)=\frac{\lambda}{4}\left(\phi^2-\mu^2\right)^2\rightarrow\frac{\lambda}{4}\,\phi^4.$$ Another very interesting choice corresponds to the power-law type of potentials: $V(\phi)=V_0\phi^{2n}$, where $n$ is a real number. In this case: $$v=\frac{2n}{\phi}\rightarrow 0,$$ so that slow-roll inflation occurs at very large field values $\phi\gg 2n$. Hence, in general, the critical point associated with slow-roll inflation is approached if initial conditions are given at very large field values, so that what we have is a chaotic inflationary scenario. 

We want to conclude this section by noting that the quadratic potential, that is usually associated with the massive scalar model, is not the only one that leads to slow-roll inflation in the framework of minimally coupled scalar field models. As a matter of fact any potential of the power-law type (positive power) leads to the existence of the slow-roll inflationary critical point $P_\text{inf}$. The same is true for the symmetry-breaking type of potential. The distinctive feature in these cases is that the slow-roll inflation occurs at large field values so that $P_\text{inf}$ can be associated with chaotic inflation. In other words, in the framework of single-field inflationary models where the scalar field is minimally coupled, chaotic inflation is quite generic in the sense that it is attained for a large class of potentials.


\section{Scalar-tensor theories of gravity in the phase space}\label{sect-stt}

In terms of the FRW metric the motion equations \eqref{action-mot-eq} for scalar-tensor theories can be written in the following way:

\bea &&3FH^2=\frac{\dot\phi^2}{2}+V-3F_\phi H\dot\phi+\rho_m,\nonumber\\
&&-2F\dot H=\left(1+F_{\phi\phi}\right)\dot\phi^2+F_\phi\left(\ddot\phi-H\dot\phi\right)+p_m+\rho_m,\nonumber\\
&&\ddot{\phi}+3H\dot\phi+\frac{F_\phi(1+3F_{\phi\phi})}{2F+3F^2_\phi}\,\dot\phi^2=\frac{4F_\phi V-2FV_\phi}{2F+3F^2_\phi}+\frac{F_\phi(\rho_m-3p_m)}{2F+3F^2_\phi},\label{mot-eq}\eea where the overdot means derivative with respect to the cosmic time $t$, while $\rho_m$ and $p_m$ are the energy density and barotropic pressure of matter, respectively. The following slow-roll parameters can be defined \cite{Carlos}:

\bea \xi_1:=-\frac{\dot H}{H^2},\;\xi_2:=\frac{\ddot\phi}{H\dot\phi},\;\xi_3:=\frac{F_\phi\dot\phi}{2FH},\;\xi_4:=\frac{E_\phi\dot\phi}{2EH},\label{slow-roll-c'}\eea with $\xi_i\ll 1$ and $\dot\xi_i\approx 0$ during inflation. Here, the quantity $E$ is defined as $E:=(2F+3F^2_\phi)/2$.

In what follows we shall focus in the search for inflationary behavior, therefore we shall omit the matter contribution.


\subsection{Dynamical System}

As it has been explained above, in general one can trade the very complex system of second order equations \eqref{mot-eq} by a system of autonomous ordinary differential equations through choosing adequate variables of some state space. Let us start by choosing the following phase space variables:

\bea x\equiv\frac{\dot\phi}{\sqrt{6F}H},\;\;y\equiv\frac{F_\phi}{\sqrt{F}},\;\;z\equiv\frac{V_\phi\sqrt{F}}{V}.\label{var}\eea In terms of these variables the Friedmann equation in \eqref{mot-eq} can be written as the following constraint:

\bea \bar\Omega_V\equiv\frac{V}{3H^2F}=1+\sqrt{6}xy-x^2.\label{friedman-c}\eea This constraint can be used to remove terms with the potential $V$ from further equations.

The equations \eqref{mot-eq} can be traded by the following dynamical system on the variables $x$, $y$ and $z$:

\bea &&x'=\frac{\ddot\phi}{\sqrt{6F}\,H^2}-\sqrt{\frac{3}{2}}x^2y-x\frac{\dot H}{H^2},\nonumber\\
&&y'=\sqrt{\frac{3}{2}}xy^2\left(2\Gamma_F-1\right),\nonumber\\
&&z'=\sqrt{\frac{3}{2}}xz\left[y+2z\left(\Gamma_V-1\right)\right],\label{edo}\eea where the prime denotes derivative with respect to the number of $e$-foldings $N=\ln a$, $\Gamma_X\equiv XX_{\phi\phi}/X^2_\phi$, and

\bea &&\frac{\ddot\phi}{\sqrt{6F}\,H^2}=\frac{-3x\left[2+\sqrt{6}xy(1+y^2\Gamma_F)-y^2\right]+\sqrt{6}(2y-z\bar\Omega_V)}{2+3y^2},\nonumber\\
&&\frac{\dot H}{H^2}=\frac{\sqrt{6}x\left[4y-\sqrt{6}x(1+y^2\Gamma_F)\right]-3y(2y-z\bar\Omega_V)}{2+3y^2}.\label{symb}\eea While deriving the equations \eqref{edo} and \eqref{symb} we have taken into account the constraint \eqref{friedman-c}.


\subsection{Generic critical points}

In general the functions $V(\phi)$ and $F(\phi)$ must be known in order to find the critical points of the 3D dynamical system (\ref{edo}). However, there are three equilibrium configurations that can be found quite independent of both $V(\phi)$ and $F(\phi)$. The corresponding critical points are the following ones:\footnote{Since neither $F(\phi)$ nor $V(\phi)$ are known we can not say anything about the stability of the given critical points.}

\begin{itemize}

\item Source points, $P_\text{stiff}:\left(\pm 1,0,0\right)$. These correspond to stiff-matter solutions since for these critical points $\bar\Omega_V=V/3H^2F\rightarrow 0$ $\Rightarrow 3FH^2\gg V$, and $$x=\pm 1\Rightarrow 3FH^2=\frac{\dot\phi^2}{2}\gg V.$$
		 
\item Inflationary manifold, ${\cal P}_\text{dS}:\left(0,z/2,z\right)$. Points in this manifold correspond to the de Sitter solution since, by substituting $x=0$ and $y=z/2$ in the second equation in \eqref{symb}, and taking into account that $x=0\Rightarrow\bar\Omega_V=1,$ it follows that $\dot H/H^2=0\Rightarrow H=H_0.$ For points in ${\cal P}_\text{dS}$ we obtain, besides, that

\bea \frac{V_\phi}{V}=2\frac{F_\phi}{F}\;\Rightarrow\;V(\phi)=V_0F^2(\phi).\label{v-f-rel}\eea This means that for the coupling function $F=\epsilon\phi^2$ the de Sitter solution exists, in particular, for potentials that at large field values $\phi\gg 1/\sqrt\epsilon$, asymptote to the quartic potential $V\propto\phi^4$. Meanwhile, for $F=1-\epsilon\phi^2$ it exists for the symmetry-breaking type of potential:\footnote{We underline that, although integration in quadratures of the condition \eqref{v-f-rel} leads to specific type of potentials, other potentials may as well fulfill this condition. In these cases \eqref{v-f-rel} determines the values of the scalar field that are to be associated with the de Sitter attractor. Below we shall come back to this issue again.} $$V\propto(\phi^2-\phi_0^2)^2,\;\phi_0=1/\sqrt\epsilon.$$  
   	
\end{itemize} The stiff-matter equilibrium configurations above exist but for the cases when $y^2(2\Gamma_F-1)$ has no zeroes. In this case the only surviving critical points are those in the inflationary manifold ${\cal P}_\text{dS}$ (see below).

We want to underline that the above conclusions are valid only in general terms. When more specific scenarios are considered, other possibilities arise, including new equilibrium configurations not included above and/or other potentials that can lead to de Sitter expansion. In the next sections we shall illustrate this statement by considering specific functional forms for the coupling function that are frequently encountered in the bibliography.


\section{Non-minimal coupling theories}\label{sect-nmc}

Non-minimal coupling theories are given by the following choice of the coupling function:

\bea F(\phi)=1-\epsilon\phi^2\Rightarrow\Gamma_F=-\frac{1-\epsilon\phi^2}{2\epsilon\phi^2}=-\frac{2\epsilon}{y^2},\label{nmc-cf}\eea where, in order to have attractive gravity $|\phi|\leq 1/\sqrt\epsilon$. When $\epsilon=1/6$ it is known as conformal coupling theory. Meanwhile, when $\epsilon=0$, we deal with the minimal coupling theories where the scalar field has not gravitational effects beyond those of a matter source.

For the above choice of the coupling function it is verified that $$y^2(2\Gamma_F-1)=-(4\epsilon+y^2)\neq 0,$$ has no zeroes. The dynamical system corresponding to the vacuum NMC theory reads:

\bea &&x'=-\sqrt{\frac{3}{2}}x^2y-x\frac{\dot H}{H^2}+\frac{\ddot\phi}{\sqrt{6F}\,H^2},\nonumber\\
&&y'=-\sqrt{\frac{3}{2}}x(y^2+4\epsilon),\nonumber\\
&&z'=\sqrt{\frac{3}{2}}xz(2\alpha z+y),\label{nmc-edo}\eea where the phase space variables are defined in \eqref{var} and $\alpha\equiv\Gamma_V-1$. Besides, 

\bea &&\frac{\dot H}{H^2}=\frac{\sqrt{6}x\left[4y-\sqrt{6}(1-2\epsilon)x\right]-3y(2y-z\bar\Omega_V)}{2+3y^2},\nonumber\\
&&\frac{\ddot\phi}{\sqrt{6F}\,H^2}=\frac{-3x\left[2+\sqrt{6}(1-2\epsilon)xy-y^2\right]+\sqrt{6}(2y-z\bar\Omega_V)}{2+3y^2},\label{nmc-symbol}\eea where $\bar\Omega_V$ is given by \eqref{friedman-c}: $\bar\Omega_V\equiv 1+\sqrt{6}xy-x^2$.

A crude inspection of the dynamical system \eqref{nmc-edo} shows that, as expected, the only equilibrium configurations belong in the inflationary de Sitter manifold:\footnote{As a matter of fact there can be other critical points at infinity, but in order to find them one needs of a complementary study.} $${\cal P}_\text{dS}:\left(0,\frac{z}{2},z\right),$$ whose critical points are attractors, as shown by the numeric investigation. Since at the de Sitter point: $$y=\frac{z}{2}\Rightarrow\frac{V_\phi}{V}=2\frac{F_\phi}{F},$$ these configurations exist, in particular, for the ''symmetry-breaking'' type of potential $$V=\frac{\lambda}{4}\,(\phi^2-\phi^2_0)^2,\;-\phi_0\leq\phi\leq\phi_0,$$ where $\phi_0\equiv 1/\sqrt\epsilon$, $\lambda\equiv 4V_0\epsilon^2$. In this case the equilibrium configuration is not associated with the minima of the symmetry-breaking potential, since at the minima, which in the present case coincide with the end points of the allowed $\phi$-interval: $\phi=\pm\phi_0$, the self-interaction potential vanishes $V=0$, which is not compatible with the de Sitter condition: $\bar\Omega_V=1$ $\Rightarrow 3H^2F=V\propto F^2$ $\Rightarrow 3H^2\propto F$, as long as at the minima of the potential the coupling function vanishes as well, $F=0$. This leads, in turn, to $H=0$, i. e., we get static universe instead of de Sitter expansion. 

Given that for the de Sitter attractor $x=0$ ($\Rightarrow\phi=$const.), which means that $\bar\Omega_V=1$, a look at the dynamical system \eqref{nmc-edo} shows that the equilibrium point exists for arbitrary potentials as long as the condition: 

\bea y=\frac{z}{2}\Rightarrow\frac{V_\phi}{V}=2\frac{F_\phi}{F}=\frac{4\epsilon\phi}{\epsilon\phi^2-1},\label{nmc-cond}\eea is fulfilled. Although integration in quadratures of \eqref{nmc-cond} straightforwardly leads to the symmetry-breaking type of potential, this is not the only type of potential that can be associated with the de Sitter critical point. Take, for instance, the well-known exponential potential $V=V_0\exp(\sigma\phi)$. In this case fulfillment of the condition \eqref{nmc-cond} is possible when, $$\phi^*_\pm=\frac{2}{\sigma}\left(1\pm\sqrt{1+\frac{\sigma^2}{4\epsilon}}\right).$$ For very small values of the non-minimal coupling $\epsilon$, as dictated by the physical evidence, the slow-roll de Sitter attractor is approached at large field values: $\phi^*_\pm\approx\pm 1/\sqrt\epsilon$, similar to the minimal coupling case discussed above. Yet another example is given by the power-law potentials: $V=V_0\phi^{2n}$, where $n$ is a real number. In this case \eqref{nmc-cond} entails that: $$\phi^*_\pm=\pm\sqrt\frac{n}{(n-2)\epsilon},$$ which is again a large quantity if $\epsilon$ is small enough. Notice that, in the particular case of the quartic potential fulfillment of the condition \eqref{nmc-cond} requires that $\epsilon\phi^2\gg 1$ which is physically dismissed since we expect gravity to be attractive. Hence, in the framework of NMC single-inflation models the quartic potential can not drive the slow-roll inflation.

Below we shall investigate the physically outstanding case of the Brans-Dicke (BD) theory. In this case a certain simplification of the dynamics is achieved so that we can perform an exhaustive study of the phase space global dynamics.


\section{Inflationary dynamics of vacuum Brans-Dicke theory}\label{sect-bd}

In this section we study the scalar-tensor gravity that is specified by the following choice of the function $F=F(\phi)$ in \eqref{mot-eq}:

\bea F(\phi)=\epsilon\phi^2,\label{bd-choice}\eea where $\epsilon$ is a dimensionless coupling constant. It can be easily checked that, under the simultaneous replacement \cite{fujii-book}:

\bea \phi\rightarrow\frac{1}{2}\,\epsilon\phi^2,\;\omega_\text{BD}\rightarrow\frac{1}{4\epsilon},\label{phi2-to-bd}\eea the theory given by the choice \eqref{bd-choice} maps into the Jordan frame (JF) BD theory, 

\bea {\cal L}^{\phi^2}_\text{BD}=\frac{1}{2}\,\epsilon\phi^2R-\frac{1}{2}(\der\phi)^2\rightarrow{\cal L}^\text{JF}_\text{BD}=\phi R-\frac{\omega_\text{BD}}{\phi}(\der\phi)^2,\label{bd-lag}\eea so that the choice \eqref{bd-choice} corresponds to choosing Brans-Dicke theory \cite{bd-1961}. The equations of motion \eqref{mot-eq} are greatly simplified in this case:

\bea &&H^2=\frac{1}{6\epsilon}\left(\frac{\dot\phi}{\phi}\right)^2-2\frac{\dot\phi}{\phi}\,H+\frac{V}{3\epsilon\phi^2}+\frac{\rho_m}{3\epsilon\phi^2},\nonumber\\
&&\dot H=-\frac{1}{2\epsilon}\left(\frac{\dot\phi}{\phi}\right)^2+4H\frac{\dot\phi}{\phi}-\frac{4V-\phi V_\phi}{(1+6\epsilon)\phi^2}-\frac{p_m+(1+8\epsilon)\rho_m}{2\epsilon(1+6\epsilon)\phi^2},\nonumber\\
&&\frac{\ddot\phi}{\phi}+3H\frac{\dot\phi}{\phi}+\left(\frac{\dot\phi}{\phi}\right)^2=\frac{4V-\phi V_\phi}{(1+6\epsilon)\phi^2}+\frac{\rho_m-3p_m}{(1+6\epsilon)\phi^2}.\label{bd-mot-eq}\eea


\subsection{Comparison of the goals of the present approach with those of similar studies} 

Before we continue with the investigation of the asymptotic dynamics of BD theory \eqref{bd-mot-eq}, we want to mention that a similar study has been performed before, among others, in references \cite{hrycyna-2013} and \cite{richard-prd-2015}. In \cite{hrycyna-2013} the dynamics of the Brans-Dicke theory in the Jordan frame was investigated. Cosmological models that are based in the following JFBD Lagrangian density:

\bea {\cal L}^\text{JF}_\text{BD}=\phi R-\frac{\omega_\text{BD}}{\phi}(\der\phi)^2-2V,\label{jfbd-lag}\eea were considered with the inclusion of the matter Lagrangian. The work \cite{hrycyna-2013} was focused in the search for global attractor dynamics representing a de Sitter state, however the authors were not interested in the whole global dynamics since not all of the existing critical points were found. As a matter of fact, at least one of the variables of the phase space chosen in \cite{hrycyna-2013}: $$x^*=\frac{\dot\phi}{H\phi},\;y^*=\sqrt\frac{V}{3\phi}\,\frac{1}{H},\;\lambda^*=-\phi\frac{V_\phi}{V},$$ the variable $\lambda^*$ to be specific, is unbounded so that, in order to find the critical points at $\lambda^*$-infinity, a complementary study -- not performed in the mentioned reference -- is required. According to the results of \cite{hrycyna-2013} there are values of the BD coupling parameter $\omega_\text{BD}$ for which a global attractor in the phase space representing the de Sitter stage exists. 

The investigation of \cite{hrycyna-2013} was revisited in \cite{richard-prd-2015}, where it was concluded that in the JFBD the de Sitter solution is found only for the quadratic potential. The authors of \cite{richard-prd-2015} considered the BD theory in the dilatonic frame (also string frame) that is given by the Lagrangian density: 

\bea {\cal L}_\text{BD}^\text{dil}=e^\vphi\left[R-\omega_\text{BD}(\der\vphi)^2\right]-2V.\label{bd-dil-lag}\eea Under the replacement $\phi=e^\vphi$, the latter Lagrangian is transformed into the JFBD Lagrangian density \eqref{jfbd-lag} above. In this work the authors considered the exponential self-interaction potential and its combinations, such as the $\cosh$ and $\sinh$ potentials (the trivial case with the constant potential was also considered). In this regard we should notice that, since the dilatonic BD theory was the considered framework, then in terms of the usual JFBD theory field variables \eqref{jfbd-lag}, the exponential and its combinations: $\cosh$ and $\sinh$ potentials, all of them amount to the power-law potential and its combinations: $$e^{k\vphi}=\phi^k,\;\cosh(k\vphi)=\frac{1}{2}\left(\phi^k+\phi^{-k}\right),\;\;\text{etc.}$$  

In the present paper we shall explore the BD theory in the formulation given by the Lagrangian density:

\bea {\cal L}^{\phi^2}_\text{BD}=\frac{\epsilon\phi^2}{2}\,R-\frac{1}{2}\,(\der\phi)^2-V,\label{f2bd-lag}\eea that, as shown, is equivalent to the JFBD Lagrangian \eqref{jfbd-lag}. Besides, we consider the exponential potential $V\propto\exp(\sigma\phi)$ which under the above replacements transforms into the exponential in the JFBD theory: $\exp(\sigma\phi)\rightarrow\exp(\sigma\sqrt{8\omega_\text{BD}\,\phi})$, and also the symmetry-breaking potential $V\propto(\phi^2-\mu^2)^2$ which transforms into $V\propto(\phi-\mu^2/8\omega_\text{BD})^2$, in the JFBD formulation. None of these potentials have been considered before in similar studies. Due to its role in the study of the inflationary dynamics, we also dedicate a separate subsection to discussing on the role of the power-law potentials $V\propto\phi^{2n}$. 

The investigation in this section differs from the ones in \cite{hrycyna-2013} and in \cite{richard-prd-2015}, in the following aspects: i) here we use a different framework for the BD theory that is evident from \eqref{f2bd-lag}, ii) but for the power-law potential here we consider self-interaction potentials that were not explored in \cite{richard-prd-2015} and also make general statements that are independent of the type of potential as in \cite{hrycyna-2013}, and iii) we use a different set of phase space variables than the ones used in the mentioned references, which allows us to expose the whole global dynamics of the phase space of the Brans-Dicke theory. Besides, we complement the study with the numeric investigation as well.


\subsection{The dynamical system} 

The cosmological equations \eqref{bd-mot-eq} can be written in terms of the following phase space variables:

\bea x\equiv\frac{\dot\phi}{\sqrt{6\epsilon}\,\phi H},\;z\equiv\sqrt\epsilon\frac{\phi V_\phi}{V},\;w\equiv\frac{\sqrt{\rho_m}}{\sqrt{3\epsilon}\,\phi H}.\label{vars-xzw}\eea In the present case, since $F=\epsilon\phi^2$, $y$ in \eqref{var} is not a variable of the phase space but it is a real number: $y=2\sqrt\epsilon$. The Friedmann equation in \eqref{bd-mot-eq} can be written as the following Friedmann constraint relating the kinetic and potential energy densities of the scalar field with the matter energy density:

\bea \bar\Omega^w_V\equiv\frac{V}{3\epsilon\phi^2H^2}=1+2\sqrt{6\epsilon}\,x-x^2-w^2.\label{fried-c}\eea The latter equation allows us to remove terms with the potential $V$ from the subsequent equations. We have that:

\bea \frac{\ddot\phi}{\phi H^2}=-3\sqrt{6\epsilon}\,x-6\epsilon x^2+\frac{3\sqrt\epsilon\left[\sqrt\epsilon(4-3\gamma_m)w^2+(4\sqrt\epsilon-z)\bar\Omega^w_V\right]}{1+6\epsilon},\label{useful-1}\eea which can be written in the following alternative form:

\bea \frac{\ddot\phi}{\dot\phi H}=-3-\sqrt{6\epsilon}\,x+\sqrt\frac{3}{2}\frac{\sqrt\epsilon(4-3\gamma_m)w^2+(4\sqrt\epsilon-z)\bar\Omega^w_V}{(1+6\epsilon)x},\label{ddotphi}\eea and

\bea \frac{\dot H}{H^2}=4\sqrt{6\epsilon}\,x-3x^2-\frac{3\left[(\gamma_m+8\epsilon)w^2+2\sqrt\epsilon(4\sqrt\epsilon-z)\bar\Omega^w_V\right]}{2(1+6\epsilon)},\label{useful-2}\eea where we have assumed the following relationship, $p_m=(\gamma_m-1)\rho_m$, between the pressure and the energy density of the matter fluid ($\gamma_m$ is the barotropic index of the fluid).

The motion equations \eqref{bd-mot-eq} can be traded by the following 3D dynamical system:

\bea &&x'=\frac{1}{\sqrt{6\epsilon}}\left(\frac{\ddot\phi}{\phi H^2}\right)-\sqrt{6\epsilon}\,x^2-x\frac{\dot H}{H^2},\nonumber\\
&&z'=\sqrt{6}\,xz\left(\alpha z+\sqrt\epsilon\right),\nonumber\\
&&w'=-w\left(\frac{3\gamma_m}{2}+\sqrt{6\epsilon}\,x+\frac{\dot H}{H^2}\right),\label{edo-xzw}\eea where, as before, the prime means derivative with respect to the time variable $N=\ln a$, and $\alpha\equiv\Gamma_V-1$. In the above equations $\ddot\phi/\phi H^2$ and $\dot H/H^2$ should be substituted from equations \eqref{useful-1} and \eqref{useful-2}, respectively. While deriving the third equation in \eqref{edo-xzw} we have taken into account the continuity equation for pressureless matter: $\dot\rho_m=-3\gamma_m H\rho_m$.

Below, for simplicity, we shall omit the matter component, i. e., in the above equations we set $w=0$. In this case the dynamical system \eqref{edo-xzw} simplifies to the following 2D dynamical system:

\bea &&x'=\sqrt\frac{3}{2}\,\bar\Omega^0_V\left[-\sqrt{6}\,x+\frac{(1+\sqrt{6\epsilon}\,x)(4\sqrt\epsilon-z)}{1+6\epsilon}\right],\nonumber\\
&&z'=\sqrt{6}\,xz(\alpha z+\sqrt{\epsilon}),\label{ode-ind-grav}\eea where 

\bea \bar\Omega^0_V\equiv\bar\Omega^{w=0}_V=1+2\sqrt{6\epsilon}\,x-x^2.\label{fried-c'}\eea Since $\bar\Omega^0_V\geq 0$ is a non-negative quantity, from \eqref{fried-c'} it follows that the variable $x$ is a bounded variable: $a_-\leq x\leq a_+$, where $a_\pm=\sqrt{6\epsilon}\pm\sqrt{6\epsilon+1}$.

The slow-roll conditions \eqref{slow-roll-c'} amount to given trajectories $z=z(x)$ in the phase plane $xz$. We have, in particular, that:

\bea &&\left|\frac{\dot H}{H^2}\right|\ll 1\Rightarrow z_*=4\sqrt\epsilon-\frac{(1+6\epsilon)x(4\sqrt{6\epsilon}-3x)}{3\sqrt\epsilon(1+2\sqrt{6\epsilon}\,x-x^2)},\label{slw-roll-xz-1}\\
&&\left|\frac{\ddot\phi}{\dot\phi H}\right|\ll 1\Rightarrow z_{**}=4\sqrt\epsilon-\frac{(1+6\epsilon)x(\sqrt{6}+2\sqrt\epsilon\,x)}{1+2\sqrt{6\epsilon}\,x-x^2}.\label{slw-roll-xz-2}\eea

Here we shall not specify the type of potential by considering arbitrary $\alpha=\alpha(x,z)$, but in order to perform the numeric investigation we shall explore very simple particular cases when $\alpha$ is either a known function of the phase space variables or a constant parameter. We shall consider, in particular, two well-known potentials of cosmological interest: the exponential potential (EXP), 

\bea V(\phi)=V_0\exp\left(\sigma\phi\right)\;\Rightarrow\;\Gamma_V=1\;\Rightarrow\;\alpha=0,\label{exp-pot}\eea where $\sigma$ is a free parameter, and the symmetry-breaking potential (SBP),

\bea V(\phi)=\frac{\lambda}{4}\left(\phi^2-\mu^2\right)^2\;\Rightarrow\;\alpha=\frac{2\sqrt\epsilon-z}{2z},\label{s-b-pot}\eea where $\mu$ and $\lambda$ are free constants. In consequence, for the EXP the second equation in \eqref{ode-ind-grav} can be written as

\bea z'=\sqrt{6\epsilon}\,xz,\label{z-exp}\eea while for the SBP we have that:

\bea z'=\sqrt\frac{3}{2}\,xz(4\sqrt\epsilon-z).\label{z-s-b}\eea


\subsection{Bounded variables: finite phase space}

The problem with the dynamical system \eqref{ode-ind-grav} is that, although $x$ is a bounded variable: $a_-\leq x\leq a_+$, the phase plane $x,z\in R^2$ is infinite in general since $z$ is unbounded. In consequence one or several critical points of the dynamical system may be located at $z$-infinities. In such a case it is recommendable to work in a different set of variables. An approach frequently used is to keep working with the unbounded variables and then to complement the study with the procedure based on the projection onto the equator of the Poincar\`e sphere -- also known as Poincar\`e compactification -- in order to bring the points at infinity into a finite region \cite{bohmer-rev, genly-plb-2007}. This was the approach followed, for instance, in the early paper \cite{zeldovich-1985}. However, there are other possible approaches as, for instance, the one exposed in \cite{uggla}, where a regular dynamical system on a bounded state space is obtained by introducing polar coordinates. A similar procedure was proposed in \cite{urena-prd-2016}. 

Here we follow a different approach that is applicable to any unbounded variables. The main idea is to introduce bounded variables $|X^\pm_i|\leq 1$ that are related with the original unbounded ones, $-\infty<\bar X_i<\infty$, in the following way \cite{quiros-cqg-2018}: $$X^+_i=\frac{\bar X_i}{1+\bar X_i},\;0\leq\bar X_i<\infty\;\left|\;X^-_i=\frac{\bar X_i}{1-\bar X_i},\;-\infty<\bar X_i\leq 0.\right.$$ 

As seen, our approach operates at the cost of introducing several sets of bounded variables since, in general, one single set of them is not enough to cover the whole phase space. In the present case we use two sets of variables: $(x,v_+)$ and $(x,v_-),$ where, in place of $z$ we introduce the new variables:

\bea v_+\equiv\frac{z}{1+z}\;(z\geq 0),\;v_-\equiv\frac{z}{1-z}\;(z\leq 0).\label{z-var}\eea  The set of variables: $a_-\leq x\leq a_+$, $0\leq v_+\leq 1$, covers the 'upper' half of the phase plane: 

\bea \Psi^+(x,z)=\{(x,z)|a_-\leq x\leq a_+,0\leq z<\infty\}\rightarrow \Psi^+(x,v_+)=\{(x,v_+)|a_-\leq x\leq a_+,0\leq v_+\leq 1\},\label{+half}\eea while a second set $x,v_-$ where $-1\leq v_-\leq 0$, covers the lower half of the phase plane:

\bea \Psi^-(x,z)=\{(x,z)|a_-\leq x\leq a_+,-\infty<z\leq 0\}\rightarrow \Psi^-(x,v_-)=\{(x,v_-)|a_-\leq x\leq a_+,-1\leq v_-\leq 0\}.\label{-half}\eea The whole phase plane is then the union: $\Psi^\text{wh}(x,u)=\Psi^+(x,v_+)\cup\Psi^-(x,v_-)$, where $u=v_+\cup v_-$.

In terms of the bounded variables \eqref{z-var} we have that

\bea \left(\frac{\dot H}{H^2}\right)_\pm=4\sqrt{6\epsilon}\,x-3x^2-\frac{3\sqrt\epsilon\left(1+2\sqrt{6\epsilon}\,x-x^2\right)\left[4\sqrt\epsilon\mp(4\sqrt\epsilon\pm1)v_\pm\right]}{(1+6\epsilon)(1\mp v_\pm)},\label{hdot-xv}\eea while the dynamical system \eqref{ode-ind-grav} is written as:

\bea &&\frac{dx}{dT_\pm}=\frac{\sqrt{3}(1+2\sqrt{6\epsilon}\,x-x^2)}{\sqrt{2}(1+6\epsilon)}\left\{-\sqrt{6}(1+6\epsilon)x(1\mp v_\pm)+(1+\sqrt{6\epsilon}\,x)\left[4\sqrt\epsilon\mp(4\sqrt\epsilon\pm 1)v_\pm\right]\right\},\nonumber\\
&&\frac{dv_\pm}{dT_\pm}=\sqrt{6}\,xv_+(1\mp v_\pm)\left[(\alpha\mp\sqrt\epsilon)v_\pm+\sqrt\epsilon\right],\label{ode-xv}\eea where we have introduced the time variables $dT_\pm=dN/(1\mp v_\pm)$, with $N=\ln a$ as in \eqref{edo} and \eqref{ode-ind-grav} above. In the sub-indexes '$\pm$' and '$\mp$' in \eqref{hdot-xv} and \eqref{ode-xv}, the upper sign is for the upper half of the phase space while the lower sign is for the lower half. For the EXP \eqref{exp-pot} the second equation in \eqref{ode-xv} reads:\footnote{Compare with equations \eqref{z-exp} for the exponential potential and \eqref{z-s-b} for the symmetry-breaking potential.}

\bea \frac{dv_\pm}{dT_\pm}=\sqrt{6\epsilon}\,xv_\pm(1\mp v_\pm)^2,\label{ode-exp-pot}\eea

while for the SBP \eqref{s-b-pot} we have that:

\bea \frac{dv_\pm}{dT_\pm}=\sqrt\frac{3}{2}\,xv_\pm(1\mp v_\pm)\left[4\sqrt\epsilon\mp(4\sqrt\epsilon\pm 1)v_\pm\right].\label{ode-s-b-pot}\eea For the power-law potential $V\propto\phi^{2n}$ ($n$ is a real number), $v_\pm=v^\pm_0$ are constants so that the second equation in \eqref{ode-xv} is an identity and the dynamical system simplifies to a single autonomous ordinary differential equation. This case will be studied in the next subsection.

\begin{figure}[t!]\begin{center}
\includegraphics[width=7cm]{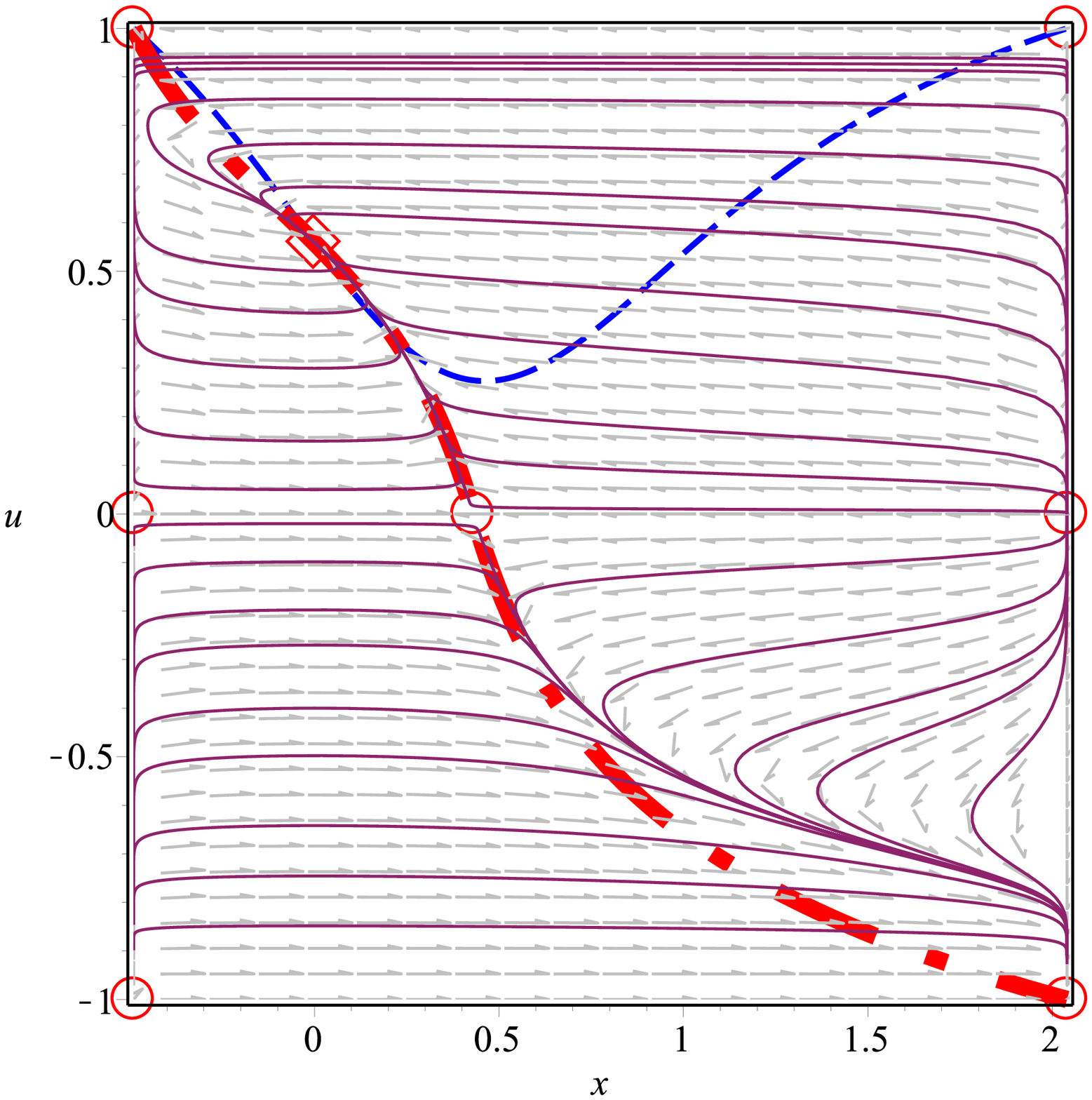}
\includegraphics[width=7cm]{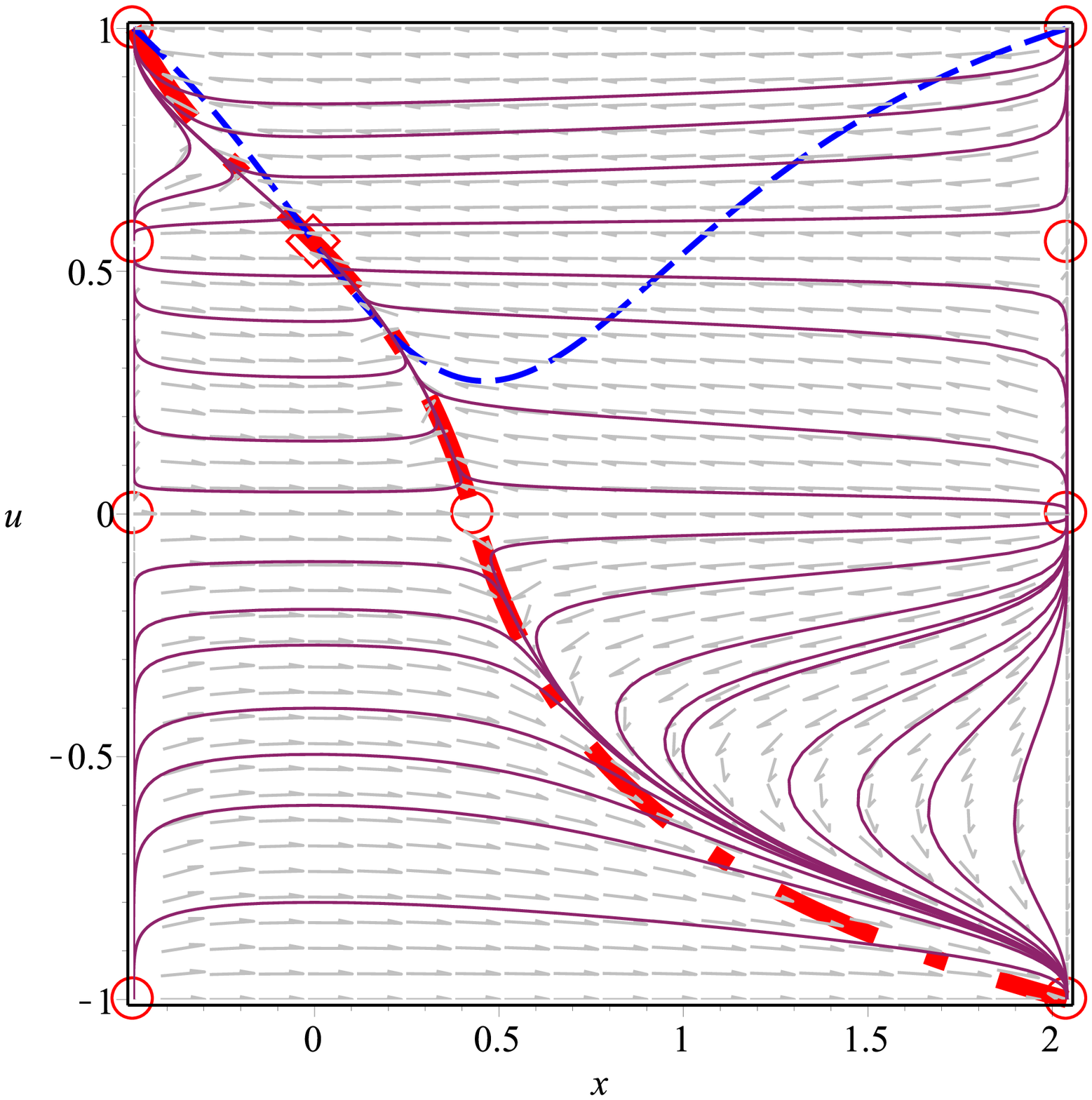}
\vspace{1.2cm}\caption{Phase portrait of the dynamical system \eqref{ode-xv} for the exponential potential \eqref{exp-pot} -- left hand panel -- and for the symmetry-breaking potential \eqref{s-b-pot} -- right hand panel. We have arbitrarily set $\epsilon=0.1$. The small circles enclose the critical points of the dynamical system but for the slow-roll de Sitter critical point, $P_\text{dS}:\left(0,4\sqrt\epsilon/(4\sqrt\epsilon+1)\right)$, that is enclosed within the small diamond. The thin dashed curve corresponds to the slow roll condition $|\dot H/H^2|\ll 1$, meanwhile the thick dash-dot curve corresponds to the slow-roll condition $|\ddot\phi/\phi H|\ll 1$. All of the slow-roll conditions, including $\dot\phi/\phi H\ll 1\Rightarrow x\approx 0$, are jointly satisfied at the local de Sitter attractor (diamond).}\label{fig3}\end{center}\end{figure}

The slow-roll conditions \eqref{slow-roll-c'} amount to:

\bea \left|\frac{\dot\phi}{\phi H}\right|\ll 1\;\Rightarrow\;x\approx 0,\;\left|\frac{\dot H}{H^2}\right|\ll 1\;\Rightarrow\;v^*_\pm\approx\frac{z_*}{1\pm z_*},\;\left|\frac{\ddot\phi}{\dot\phi H}\right|\ll 1\;\Rightarrow\;v^{**}_\pm\approx\frac{z_{**}}{1\pm z_{**}},\label{slow-roll-uv-1}\eea where $z_*$ and $z_{**}$ are given by equations \eqref{slw-roll-xz-1} and \eqref{slw-roll-xz-2}, respectively. By looking at FIG. \ref{fig3} one may notice that the curve, $\lambda(x)=v^{**}_+\cup v^{**}_-$, approaches to the separatrix in $\Psi^\text{wh}(x,u)$ -- at least -- along of the heteroclinic orbit joining the de Sitter critical point $P_\text{dS}$ with the scaling saddle point $P_\text{sc}$ (see below).


The critical points $P_i:(x_i,u_i)$ of the dynamical system \eqref{ode-xv} in $\Psi^\text{wh}(x,u)$, are the following ones (see FIG. \ref{fig3}):

\begin{itemize}

\item Stiff-matter solutions, $$P^\pm_\text{stiff}:\left(a_\pm,0\right),$$ where $a_\pm=\sqrt{6\epsilon}\pm\sqrt{6\epsilon+1}$. The point $P^+_\text{stiff}$ is a source point while $P^-_\text{stiff}$ is a saddle critical point. In this case: $$x=a_\pm\Rightarrow\phi H=\frac{\dot\phi}{a_\pm\sqrt{6\epsilon}},\;\frac{V}{3\epsilon\phi^2H^2}\rightarrow 0\Rightarrow\phi H\gg\sqrt{V/3\epsilon}.$$ Hence, $\dot\phi/\sqrt{2}\gg a_\pm\sqrt{V}$. For the symmetry-breaking potential \eqref{s-b-pot}, due to the condition, $$v_\pm=0\Rightarrow\frac{\phi V_\phi}{V}\rightarrow 0,$$ these solutions are associated with the maximum of the potential. Actually, in this case: $$\frac{\phi V_\phi}{V}=\frac{4\phi^2}{\phi^2-\mu^2}\rightarrow 0\;\Rightarrow\;\phi\rightarrow 0.$$ Since at the minimum $V=\lambda\mu^4/4$, then: $\dot\phi\gg a_\pm\sqrt{\lambda/2}\mu^2$.

For the EXP \eqref{exp-pot}, the condition that $\phi V_\phi/V\rightarrow 0$, implies that these critical points are approached at very small field values: $\phi\ll 1/\sigma$. For the power-law potential $V\propto\phi^{2n}$, since $\phi V_\phi/V=2n$, the stiff-matter solutions $P^\pm_\text{stiff}$ do not exist.

\item Special stiff-matter solutions, $$P^\pm_\text{s-stiff}:\left(a_\pm,\frac{4\sqrt\epsilon}{4\sqrt\epsilon+1}\right),$$ which exist only for potentials that asymptotically approach to the quartic potential such as, for instance, the SBP. Actually, in this case: $$z=4\sqrt\epsilon\;\Rightarrow\;\frac{\phi V_\phi}{V}=4\;\Rightarrow\;V\propto\phi^4,$$ and, since at this point $V=0$, it is necessarily linked with the minimum of the quartic potential. The point $P^-_\text{s-stiff}$ is a source point while $P^+_\text{s-stiff}$ is a saddle instead. 

For other potentials the point does not exist. Take, for instance, the power-law potential $V=V_0\phi^{2n}$. We get that $z=4\sqrt\epsilon\;\Rightarrow\;2n=4$, so that only for $n=2$ -- the quartic potential -- the critical point exists. Meanwhile, for the exponential potential: $V=V_0\exp(\sigma\phi)$, we have that $$z=4\sqrt\epsilon\;\Rightarrow\;\sigma\phi=4\;\Rightarrow\;V=V_0\,e^4\neq 0,$$ i. e., the condition $V=0$ is never fulfilled.

\item Slow-roll de Sitter solution, $$P_\text{dS}:\left(0,\frac{4\sqrt\epsilon}{4\sqrt\epsilon+1}\right).$$ It can be either a local attractor or a saddle point. From FIG. \ref{fig3} it is seen that for the EXP the de Sitter point is always a local attractor. Meanwhile, for the SBP the slow-roll inflationary solution can be either a local attractor for initial conditions obeying $0\leq u\leq 4\sqrt\epsilon/(4\sqrt\epsilon+1)$, or a saddle point for initial conditions fulfilling $4\sqrt\epsilon/(4\sqrt\epsilon+1)\leq u\leq 1$. For initial conditions in the lower half of the phase plane ($u\leq 0$) the de Sitter critical point may not be approached.

Given that $x=0\;\Rightarrow\;\phi=\phi_0,$ and, at the same time, $$\frac{V}{3\epsilon\phi^2H^2}=1,\;\frac{\phi V_\phi}{V}=4\;\Rightarrow\;V\propto\phi^4,$$ for the BD theory the de Sitter critical point $P_\text{dS}$ exists, in particular, for potentials that asymptotically approach to the quartic potential. We want to underline that, as it has been discussed in section \ref{sect-nmc} for the NMC theory, for the BD theory the de Sitter point also exists for other potentials beyond the quartic one. Take, for instance, the EXP: $V=V_0\exp(\sigma\phi)$. In this case the condition $z=4\sqrt\epsilon$ means that, $\sigma\phi=4$. Hence, for the EXP, at the de Sitter point $$H=\frac{\sigma}{4}\sqrt\frac{V_0}{3\epsilon}\,e^2.$$ Another example can be the symmetry-breaking potential \eqref{s-b-pot}. For this potential the de Sitter solution is approached at constant $\phi_0\gg\mu$ $\Rightarrow\;V\propto\phi_0^4$, and the Friedmann constraint leads to:\footnote{Throughout the paper we consider expanding cosmologies exclusively, so that we take into account only non-negative $H\geq 0$.} $$H^2=\frac{\lambda\phi_0^2}{12\epsilon}\;\Rightarrow\;H=H_0=\frac{\phi_0}{2}\sqrt\frac{\lambda}{3\epsilon}.$$ Hence, as in the minimally coupled and NMC theories, for the SBP the de Sitter equilibrium configuration is attained at large field values $\phi_0\gg\mu$.

\item Scaling between the kinetic and potential energy densities (saddle point), $$P_\text{sc}:\left(\frac{2\sqrt{6\epsilon}}{3(2\epsilon+1)},0\right).$$ For this equilibrium configuration we get that both, 

\bea \frac{\dot\phi}{\sqrt{6\epsilon}\,\phi H}=\frac{2\sqrt{6\epsilon}}{3(2\epsilon+1)},\;\frac{V}{3\epsilon\phi^2H^2}=\frac{60\epsilon^2+28\epsilon+3}{3(2\epsilon+1)^2},\nonumber\eea are constants, so that the kinetic and the potential energies are related by a constant: $$\frac{\dot\phi^2}{2V}=\frac{8\epsilon}{60\epsilon^2+28\epsilon+3}.$$ The Hubble parameter evolves like, $$H=\frac{2\epsilon+1}{4\epsilon(t-t_0)},\;t_0=\frac{2\epsilon+1}{4\epsilon}\,C_0,$$ where $C_0$ is an integration constant. Given that at $P_\text{sc}$, $\phi V_\phi/V\rightarrow 0$, the scaling solution does not exist for the power-law potentials.

\item Generic stiff-matter solutions, $$P^\pm_{\text{g-stiff},\pm}:\left(a_\pm,\pm 1\right),$$ for which $\phi V_\phi/V\rightarrow\infty.$  For the SBP \eqref{s-b-pot} these solutions are associated with the minima of the potential since, $$\frac{\phi V_\phi}{V}=\frac{4\phi^2}{\phi^2-\mu^2}\rightarrow\infty\;\Rightarrow\;\phi\rightarrow\pm\mu.$$ Meanwhile, for the exponential potential \eqref{exp-pot}, since: $$\frac{\phi V_\phi}{V}=\sigma\phi\rightarrow\infty,$$ the points $P^\pm_{\text{g-stiff},\pm}$ are asymptotically approached a very large values of the field: $\phi\gg 1/\sigma$. For the power-law potential these solutions do not exist.

The point $P^+_{\text{g-stiff},+}:(a_+,1)$ is a saddle for the EXP and a source critical point for the SBP, meanwhile $P^-_{\text{g-stiff},+}:(a_-,1)$ is a saddle for the EXP and a local attractor for the SBP. The points $P^+_{\text{g-stiff},-}:(a_+,-1)$ and $P^-_{\text{g-stiff},-}:(a_-,-1)$ are a local attractor and a source critical points, respectively, for both potentials.

\end{itemize} 

As seen, but for the special stiff-matter solutions, $P^\pm_\text{s-stiff}$ that are found only for the SBP potential \eqref{s-b-pot}, the remaining critical points are common to both potentials \eqref{exp-pot} and \eqref{s-b-pot}. Notice, in particular, that the de Sitter solution arises for any potential that leads to $\alpha$ being a function of the phase space variable $z$: $\alpha=\alpha(z)$, including the cases when $\alpha$ is a constant. This can be seen by a simple inspection of \eqref{ode-ind-grav}, or, in terms of the compact variables: $$v'_\pm=\sqrt{6}\,xv_\pm\left[(\alpha\mp\sqrt\epsilon)v_\pm+\sqrt\epsilon\right],$$ i. e., the de Sitter critical point ($x=0$) exists for any $\alpha=\alpha(v_\pm)$. We recall, however, that not every potential admits writing $\alpha=\Gamma_V-1$ as a function of $v_\pm$ (or of $z$).


\subsection{Power-law potential}\label{subsect-quartic}

Given its singular properties, here we dedicate a particular space to study the power-law potential in the BD theory. For this potential we have that:

\bea V=V_0\phi^{2n}\;\Rightarrow\;V_\phi=2nV\phi^{-1},\label{p-law-pot}\eea where $n$ is a free constant, so that $z=2n\sqrt\epsilon$ is not a variable. For the vacuum case the dynamical system \eqref{ode-ind-grav} reduces to a single ordinary differential equation:

\bea x'=\sqrt\frac{3}{2}\left\{\frac{[2(n+1)\epsilon-1]\sqrt{6}\,x+2(2-n)\sqrt\epsilon}{1+6\epsilon}\right\}\bar\Omega_V,\label{ode-p-law}\eea where, according to \eqref{fried-c}, $$\bar\Omega_V\equiv\frac{V}{3\epsilon\phi^2H^2}=1+2\sqrt{6\epsilon}\,x-x^2.$$ As seen from \eqref{ode-p-law}, only for the quartic potential ($n=2$) the de Sitter critical point exists. This potential corresponds to the quadratic potential in the formulation of BD theory given by the Lagrangian ${\cal L}_\text{BD}$ in \eqref{bd-lag}.

\begin{figure}[t!]\begin{center}
\includegraphics[width=7cm]{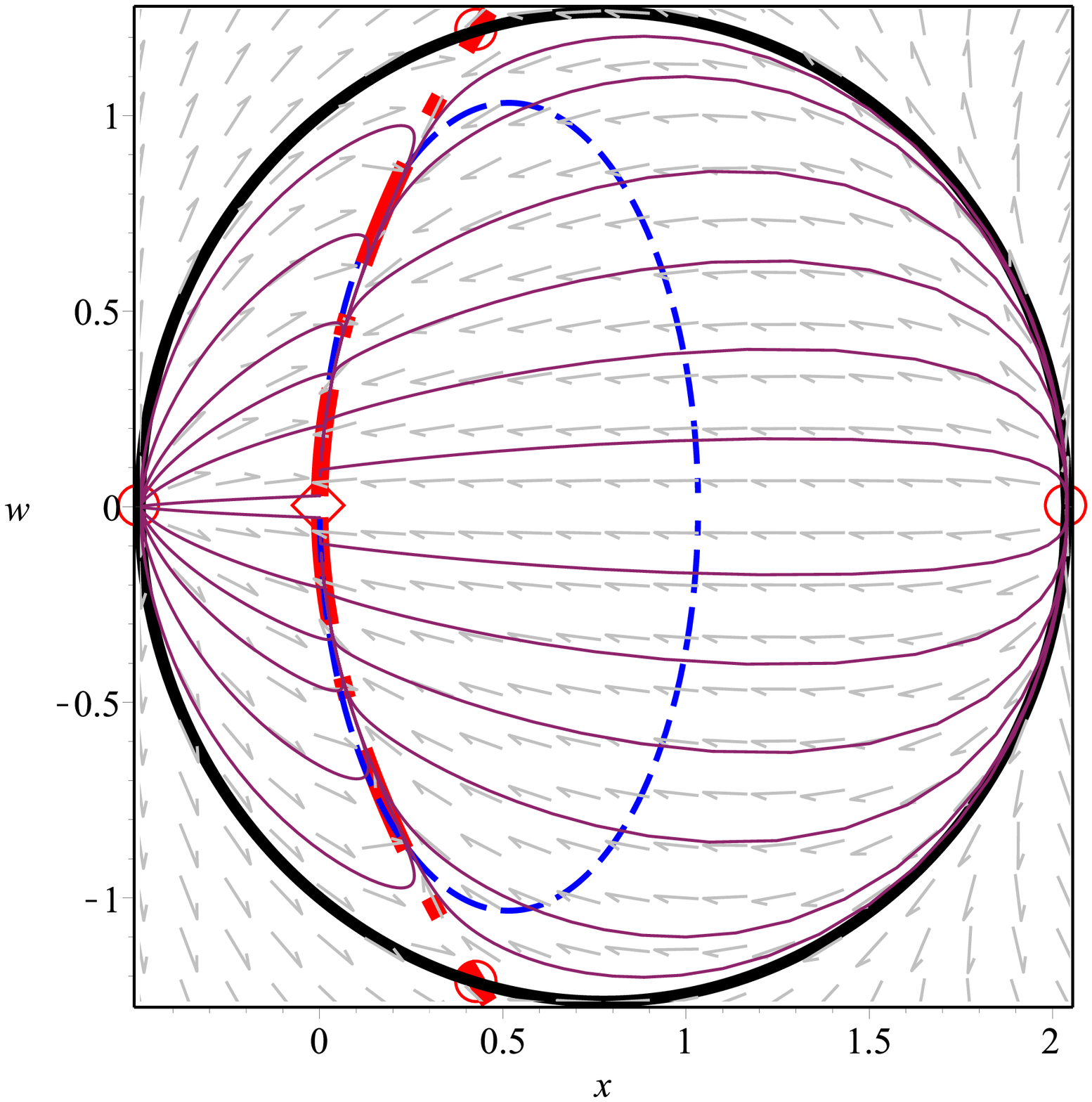}
\includegraphics[width=7cm]{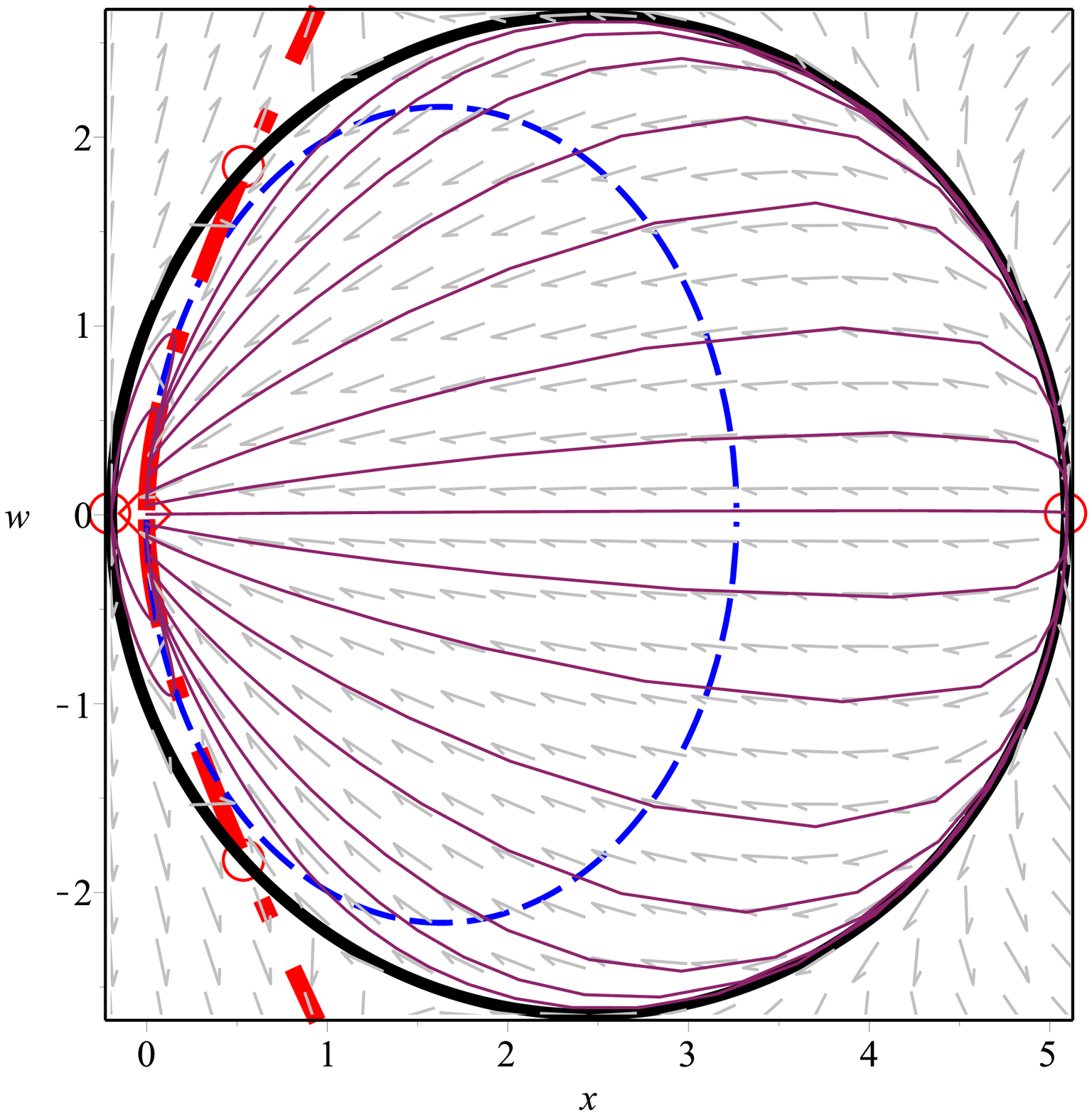}
\vspace{1.2cm}\caption{Phase portrait of the dynamical system \eqref{ode-p-law'} for the quartic potential $V\propto\phi^4$ for two different values of the free parameter: $\epsilon=0.1$ -- left hand panel -- and $\epsilon=1$ -- right hand panel. The small circles enclose the critical points of the dynamical system but for the de Sitter attractor, $P_\text{dS}:\left(0,0\right)$, that is enclosed within the small diamond. The thin dashed curve corresponds to the slow roll condition $|\dot H/H^2|\ll 1$, meanwhile the thick dash-dot curve corresponds to the slow-roll condition $|\ddot\phi/\phi H|\ll 1$.}\label{fig4}\end{center}\end{figure}

In order to check the role of the de Sitter critical point in the case of the quartic potential: $V=V_0\phi^4$, it is instructive to add a cosmological constant term $\Lambda$ into the cosmological equations. This corresponds to setting $\gamma_m=0$ and $z=4\sqrt\epsilon$ in equations \eqref{fried-c}-\eqref{edo-xzw}. We end up with the following plane-autonomous dynamical system:

\bea &&x'=-3x(1+2\sqrt{6\epsilon}\,x-x^2-w^2)+\frac{[2\sqrt{6\epsilon}-3(1+2\epsilon)x]w^2}{1+6\epsilon},\nonumber\\
&&w'=-w\left(5\sqrt{6\epsilon}\,x-3x^2-\frac{12\epsilon w^2}{1+6\epsilon}\right),\label{ode-p-law'}\eea where according to \eqref{vars-xzw} the variable $w=\sqrt\Lambda/\sqrt{3\epsilon}\phi H$ and, as before, the prime denotes derivative with respect to the time variable $N=\ln a$. The critical points of \eqref{ode-p-law'} are located within the disk:

\bea \Psi=\left\{(x,w)|a_-\leq x\leq a_+,\left(x-\sqrt{6\epsilon}\right)^2+w^2\leq 1+6\epsilon\right\},\;a_\pm=\sqrt{6\epsilon}\pm\sqrt{6\epsilon+1}.\label{ph-port-p-law}\eea The slow-roll conditions \eqref{slow-roll-c'} correspond to the following curves:

\bea &&\left|\frac{\dot H}{H^2}\right|\ll 1\;\Rightarrow\;w^*_\pm\approx\pm\sqrt{\frac{1+6\epsilon}{12\epsilon}\,x(4\sqrt{6\epsilon}-3x)},\nonumber\\
&&\left|\frac{\ddot\phi}{\dot\phi H}\right|\ll 1\;\Rightarrow\;w^{**}_\pm\approx\pm\sqrt{\frac{1+6\epsilon}{2\sqrt{6\epsilon}}\,x(3+\sqrt{6\epsilon}\,x)}.\label{slow-r-q-pot}\eea From FIG. \ref{fig4} it is seen that the curve $\lambda=\lambda(x)=\left\{w^{**}_+\cup w^{**}_-\right\}$ (thick dash-dot curve), coincides with the separatrices in $\Psi$.


Below we list the critical points of \eqref{ode-p-law'} and make a few comments on their properties.

\begin{itemize}

\item de Sitter attractor, $P_\text{dS}:(0,0)$. At this equilibrium configuration we have that $$\{x=0\;(\phi=\phi_0),\;w=0\}\;\Rightarrow\;\bar\Omega^w_V=1\;\Rightarrow\;H=H_0=\phi_0^2\sqrt\frac{V_0}{3\epsilon}.$$                         

\item Stiff-mater solutions, $P^\pm_\text{stiff}:\left(a_\pm,0\right),$ for which $$\frac{\dot\phi}{a_\pm\sqrt{6\epsilon}}=\phi H\gg V.$$ These are source critical points in $\Psi$.

\item Scaling solutions, $$P^\pm_{\text{sc},*}:\left(\frac{2\sqrt{6\epsilon}}{3(1+2\epsilon)},\pm\frac{\sqrt{180e^2+84e+9}}{3(1+2\epsilon)}\right),$$ which are saddle points in the phase disk. These represent scaling of the type: $\dot\phi/\sqrt{2\Lambda}=$const., so that $\phi(t)\propto t$. Notice that this scaling solution has nothing to do with the scaling point $P_\text{sc}$ discussed above, since $P^\pm_{\text{sc},*}$, representing scaling between the kinetic energy of the scalar field and the energy density of vacuum $\Lambda$, can not be found for the dynamical system \eqref{ode-xv} since in that case the cosmological constant was not considered.

\end{itemize}

An interesting property of the global dynamics of the BD-theory in the form \eqref{f2bd-lag}, with the quartic potential $V=V_0\phi^4$ and a cosmological constant $\Lambda$, is that unlike in the minimal coupling case where the slow-roll inflation is a saddle critical point and the $\Lambda$-de Sitter solution is the global attractor, in the present case there is not any critical point that can be associated with $\Lambda$-dominated energy density and the slow-roll inflation persists as the global attractor. This may be due to the fact that the dynamics of the scalar field is determined not by the potential $V$ but by an effective potential $W_\text{eff}$ such that: $$\frac{dW_\text{eff}}{d\vphi}=\frac{4}{1+6\epsilon}\left[\vphi\frac{dV}{d\vphi}-2\left(V+\Lambda\right)\right],$$ where we have introduced the scalar field variable $\vphi=\phi^2$. Actually, after the latter replacement of the scalar field the BDKG equation -- third equation in \eqref{bd-mot-eq} -- can be written in minimal-coupling form:

\bea \ddot\vphi+3H\dot\vphi=-\frac{4}{1+6\epsilon}\left(\vphi\frac{dV}{d\vphi}-2V\right).\label{bdkg-eq}\eea Then we introduce the effective potential $W_\text{eff}$ such that:

\bea \frac{dW_\text{eff}}{d\vphi}=\frac{4}{1+6\epsilon}\left(\vphi\frac{dV}{d\vphi}-2V\right)\;\Rightarrow\;W_\text{eff}=\frac{4}{1+6\epsilon}\int\left(\vphi\frac{dV}{d\vphi}-2V\right)d\vphi+W_0,\label{eff-pot}\eea where $W_0$ is an integration constant. It is chosen so that the effective potential be always a non-negative quantity. For the exponential potential: $V=V_0\exp(\sigma\phi)=V_0\exp(\sigma\sqrt\vphi),$ we get that: $$W_\text{eff}=\frac{4V_0\,e^{\sigma\sqrt\vphi}}{\sigma^2(1+6\epsilon)}\,\left(\sigma^2\vphi-6\sigma\sqrt\vphi+6\right)+W_0.$$ The minimum of this effective potential is at $\phi=\sqrt\vphi=4/\sigma$, where: $W^\text{min}_\text{eff}=-8V_0\,e^4/\sigma^2(1+6\epsilon)+W_0$, so that we set $W_0=8V_0\,e^4/\sigma^2(1+6\epsilon)$. In terms of the original variable $\phi$ the resulting effective potential reads:

\bea W_\text{eff}=W_\text{eff}(\phi)=\frac{4V_0\,e^{\sigma\phi}}{\sigma^2(1+6\epsilon)}\,\left(\sigma^2\phi^2-6\sigma\phi+6\right)+\frac{8V_0\,e^4}{\sigma^2(1+6\epsilon)}.\label{eff-exp-pot}\eea Hence, for the exponential potential the de Sitter attractor is associated with the minimum of this effective potential at $\phi=4/\sigma$.


The addition of a cosmological constant does not appreciably modify the main properties of the effective potential. In the present case we are interested in the quartic potential $V=V_0\phi^4$, so that integration in quadratures of the equation above for this potential yields: $$W_\text{eff}=-\frac{8V_0\Lambda\,\vphi}{1+6\epsilon}=-\frac{8V_0\Lambda\,\phi^2}{1+6\epsilon}.$$ As seen, the effective potential is a maximum at the same value $\phi=0$ at which the quartic potential is a minimum. This is why the stiff-matter solutions -- the ones associated with the maximum of the effective potential -- represent source points in the phase space, while solutions with larger values of the field are preferred.


\section{Discussion}\label{sect-discuss}

There is an ongoing discussion in the bibliography on how natural primordial inflation really is in the framework of single-field inflation models \cite{inic-infl-1, inic-infl-2, inic-infl-3, inic-infl-4, inic-infl-5}. In this regard, in \cite{inic-infl-5} an interesting statement is made that ``local dynamical systems analysis seems to show that the initial conditions for inflation do not have to be finely tuned.'' Perhaps the author refers to the kind of results discussed in \cite{zeldovich-1985, urena-ijmpd-2009} and/or in \cite{carroll-prd-2013}, even if these references were not included in \cite{inic-infl-5}. The statement is based in the known (also misleading) result that although the inflationary slow-roll trajectory is not a global attractor, it is a local attractor in initial condition space. 

Our results challenge the above statements, at least in what regards to minimal-coupling models such as the massive scalar or $\phi^2$-inflation model \eqref{phi2-action}. In this paper we have demonstrated that single-field slow-roll inflation is quite a generic stage of the expansion in the sense that the corresponding critical point exists for a large class of potentials (this includes power-law potentials with any non-negative power) and a variety of couplings of the scalar field to the curvature. Nevertheless, at least for minimal-coupling theories $\phi^2$-inflation is not as a general property of single-field inflation models as thought, since it is rather depending of the initial conditions. 

Below we shall discuss on the relative probability of inflation through identifying a rough quantitative measure for estimating the relative amount of initial data leading to slow-roll inflation. Our measure will be based in geometric probability. Imagine the phase space has a finite volume. Picking a specific point in the phase volume amounts to choosing a specific initial condition which, in turn, selects a specific orbit in the phase space. The question is: which is the probability that a given initial condition leads the related orbit to approach close enough to a given critical point? The answer is trivial if the critical point were a global attractor. In this case every possible initial condition, i. e., any point in the volume lies on an orbit that ends up in the global attractor. Given that the geometric probability is defined as the ratio of the volume containing successful initial conditions $V_\text{success}$ to the volume of the whole phase space $V_\text{whole}$: 

\bea RP=\frac{V_\text{success}}{V_\text{whole}}\times\,100\,\%,\label{rp-def}\eea the geometric probability of the global attractor is unity, i. e., $RP=100\,\%$. Meanwhile, if the critical point were a local attractor not every chosen initial condition leads to the attractor. In such a case if one were able to determine the volume of that subset of the phase space containing all of the possible initial conditions that lead to the local attractor, then one were able to give an estimate of the relative (geometric) probability by taking the ratio of the volume containing successful initial conditions to the volume of the whole phase space as in \eqref{rp-def}. Perhaps the more difficult task is to determine the relative probability of a saddle critical point. In this case one has to determine first how close to the critical point given orbits have to approach to meet appropriate physical criteria. I. e., one has to determine first a $\rho$-ball around of the saddle critical point, where the radius $\rho$ of the ball is determined through the physical criteria. Then, if one were able to determine the volume of the subset of the phase space containing all of the possible initial conditions that pick up orbits that hit the $\rho$-ball, one were able to compute a relative probability by taking the ratio of that volume to the volume of the whole phase space, just as in \eqref{rp-def}. 

Notice that the present relative probability can be implemented only if the phase space is finite, i. e., if adequate variables of the phase space are chosen. This is the case for the minimally-coupled $\phi^2$-inflation model of section \ref{sect-phi2-infl} and for BD models of inflation with arbitrary potentials in section \ref{sect-bd}. In these cases the phase space is 2D so that the geometrical probability is the ratio of areas instead of volumes.


\begin{figure}[t!]\begin{center}
\includegraphics[width=7cm]{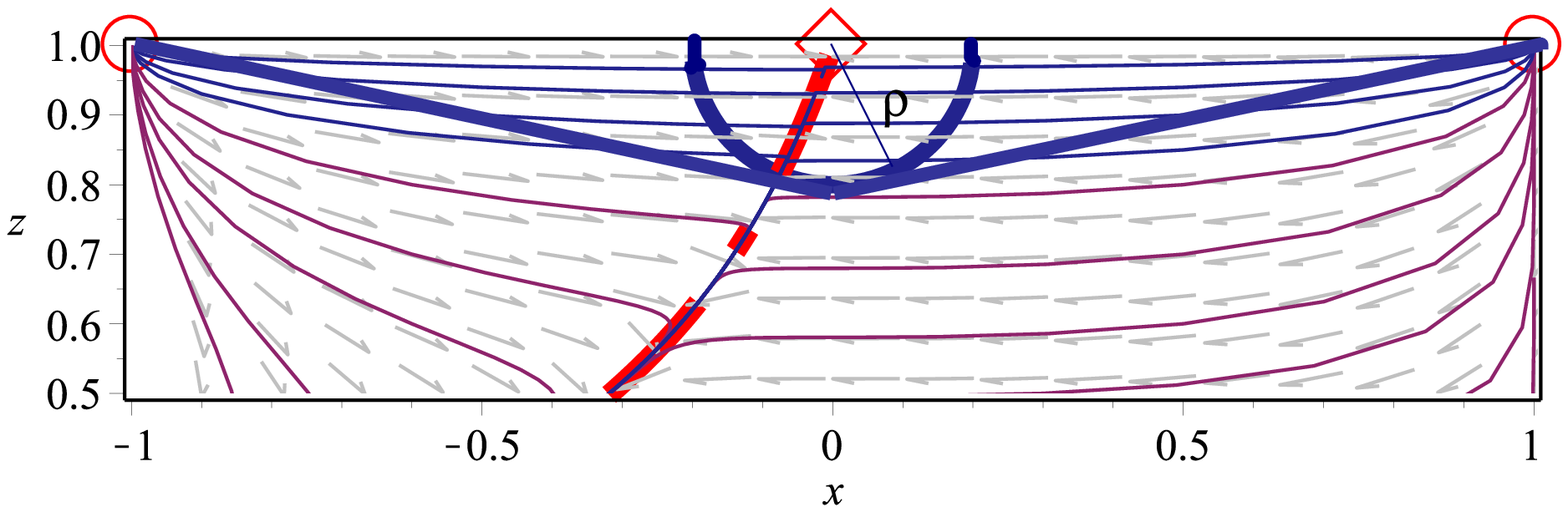}
\includegraphics[width=7cm]{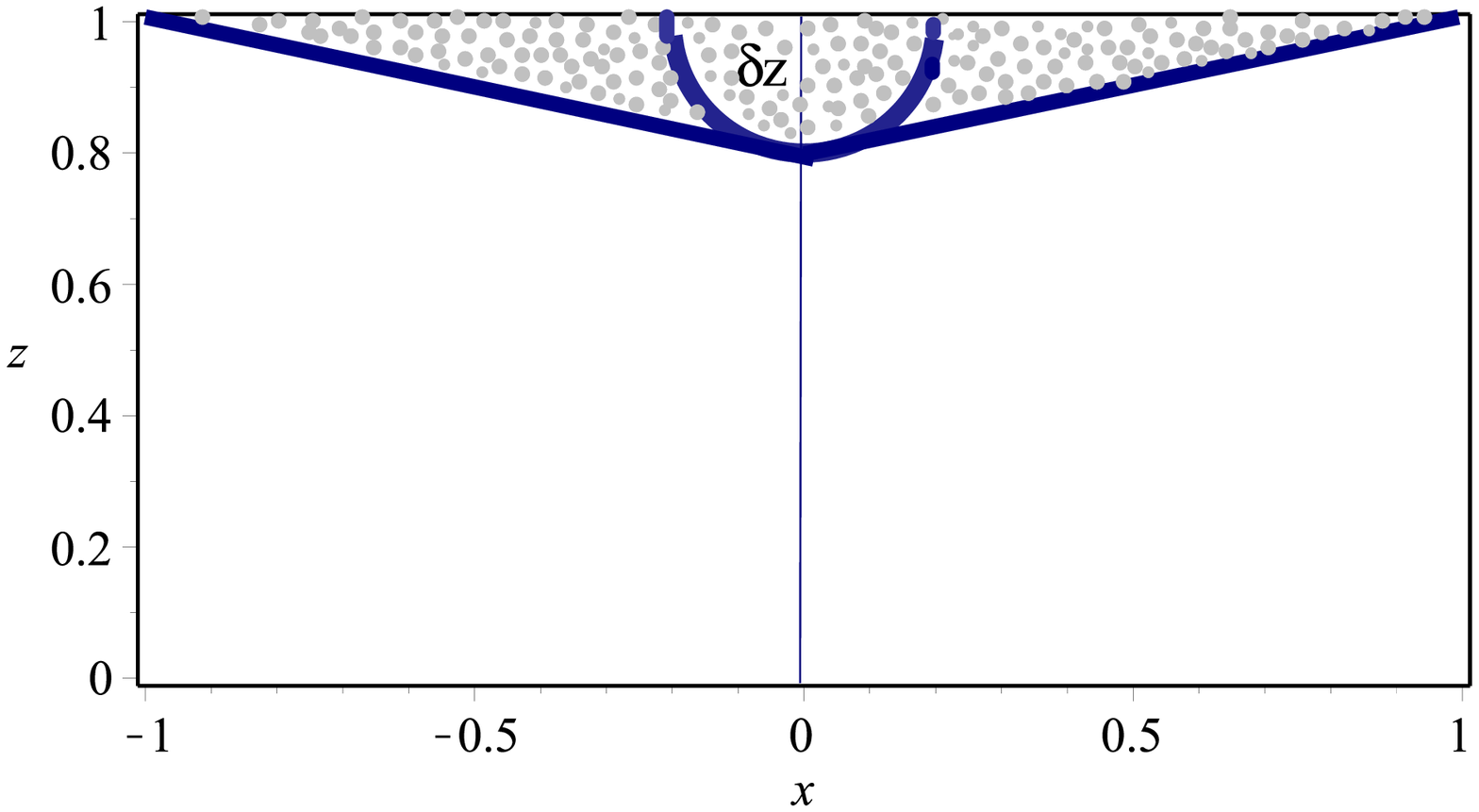}
\vspace{1cm}\caption{Towards estimating the relative amount of initial data leading to slow-roll inflation in the minimal coupling $\phi^2$-inflation model of section \ref{sect-phi2-infl}. In the left hand panel a strip of the positive branch phase portrait in FIG. \ref{fig1}, containing the de Sitter saddle point (diamond) and the $\rho$-ball around it, is shown. A triangle with vertices at $(-1,1)$, $(0,1-|\delta z|)=(0,0.8)$ and at $(1,1)$ has been drawn. Notice that any initial condition in the triangle picks up an orbit that hits the $\rho$-ball around of the slow-roll inflation saddle point. In the right hand panel the whole phase rectangle with the triangle in it (dotted area), is shown. The phase space orbits and the curves corresponding to the slow-roll conditions have been conveniently removed. Looking at the right hand panel of the figure it is seen that the geometric probability may be defined as the ratio of the area of the triangle (dotted area), to the area enclosed by the whole phase rectangle. Given that the area of the triangle is $|\delta z|\times 2/2=|\delta z|$ (the width of the rectangle is 2) and that the area of the whole phase rectangle equals $2$ (the height of the phase rectangle is unity), the ratio equals $|\delta z|/2$. In the figure, for definiteness we have arbitrarily set $|\delta z|=0.2$, however, this quantity is to be estimated from physical considerations. In this example $10\%$ of the initial data leads to slow-roll inflation. Although there is a small amount of initial data leading to inflation that falls outside of the defined triangle (see the orbits that hit the $\rho$-ball but do not lie completely within the triangle) nevertheless this does not affect the estimates.}\label{fig5}\end{center}\end{figure}



\subsection{Towards an estimation of the relative probability of inflation}

We start our analysis with the minimal-coupling $\phi^2$-inflation model (see section \ref{sect-phi2-infl}). The phase portrait corresponding to this case is given in FIG. \ref{fig1} (see also FIG. \ref{fig5} where only the positive branch of the dynamical system is considered). Notice that in this case the slow-roll heteroclinic trajectory (thick dash-dot curves in FIG. \ref{fig1} and in the left hand panel of FIG. \ref{fig5}): $$|\ddot\phi|\ll H|\dot\phi|\rightarrow 3H\dot\phi\simeq-m^2\phi\rightarrow z_\pm\simeq\frac{\sqrt{1-x^2}}{\sqrt{1-x^2}\mp 3x},$$ where we use the variables \eqref{ps-vars}, is not actually an orbit of the phase space representing a potential cosmic history. Besides, there is yet another slow-roll condition \eqref{xyz-slow-roll-c2}: $$\dot\phi^2\ll V\rightarrow 3H^2\simeq\frac{1}{2}\,m^2\phi^2\rightarrow y^2\simeq 1\rightarrow x\simeq 0,$$ that is to be fulfilled if primordial inflation is expected to take place. Both slow-roll conditions above coincide in the neighborhood of the saddle point $P_\text{inf}:(0,1)$, the one enclosed by the diamond in the figures. This neighborhood is represented by the $\rho$-ball around of the slow-roll inflationary critical point $P_\text{inf}$ (thick circle enclosing the diamond in the figure). It is defined as a ball of radius $\rho\ll 1$ (not to scale in the figure) around $P_\text{inf}$, so that orbits that hit the $\rho$-ball stay in the neighborhood of the inflationary point for enough time as to produce the $N\approx 60$ e-foldings of inflation required by the observational evidence. The radius of the ball determines the size of the set of initial conditions giving rise to the correct amount of inflation.

As explained above, in order to give a quantitative estimate of the amount of initial conditions that lead to slow-roll inflation here we choose geometric probability. In FIG. \ref{fig5} we illustrate how the geometric probability may be computed in order to obtain an order of magnitude estimate. We draw a triangle with vertexes at the source critical points $(-1,1)$, $(1,1)$ and with the third vertex at $(0,1-|\delta z|)$, where $|\delta z|=\rho$ coincides with the radius of the $\rho$-ball around of the saddle inflationary point $P_\text{inf}$. It is seen from the figure (left hand panel) that a part of several of the orbits that hit the $\rho$-ball, i. e., a small amount of initial data leading to the inflationary critical point, falls outside of this triangle. But, as we shall see, this does not affect the estimates. The area of the triangle equals $|\delta z|$, while the area of the whole phase plane is $2$ (see the right hand panel of FIG. \ref{fig5}). Hence the geometric probability is $|\delta z|/2$ and the relative probability of the initial conditions for slow-roll inflation \eqref{rp-def}: $RP\approx|\delta z|/2\times 100\,\%$. In the figure it was arbitrarily chosen the value $|\delta z|=0.2$, so that $10\%$ is the relative percent of initial data that leads to slow-roll inflation in this unphysical situation. Notice that, if in place of the chosen triangle in order to compute the geometric probability, choose a strip of height $|\delta z|=\rho$ (in which case there is not a loss of successful initial conditions, quite the contrary), we would have: $RP\approx\rho\times 100\,\%$, so that the relative probability of inflation is doubled. Yet, since the radius of the ball $\rho\ll 1$, this increase in the relative probability is negligible.

In order to give physically motivated estimates, let us to write (see the definition of the variable $z$ in \eqref{ps-vars}): $$\delta z=\frac{m\delta H}{(H+m)^2}=\frac{m\dot H}{(H+m)^2}\,\delta t.$$ This equation can be put into a simpler form if take into account the motion equations \eqref{feq} in the slow-roll approximation: 

\bea H^2=\frac{4\pi}{3M^2_\text{Pl}}\,m^2\phi^2,\;\;\dot H=-\frac{m^2}{3},\;\;\dot\phi=-\frac{m^2}{3H}\,\phi,\label{feq-slr-app}\eea where we returned to standard units ($1\rightarrow M^2_\text{Pl}/8\pi$). We obtain: $$\delta z=-\frac{m^3\delta t}{3(H+m)^2},$$ or, if take into account typical initial conditions for inflation at very large field values \cite{infl-linde-21}

\bea \frac{M_\text{Pl}}{\sqrt{12\pi}}\ll\phi\lesssim\sqrt\frac{3}{4\pi}\frac{M^2_\text{Pl}}{m},\label{phi0-bonds}\eea and $H\lesssim M_\text{Pl}$ \cite{inic-infl-3}, then: 

\bea \delta z\approx-\frac{m^3}{3M^2_\text{Pl}}\,\delta t.\label{dz}\eea Besides, $$\dot\phi=-\frac{m^2}{3H}\,\phi=-\frac{mM_\text{Pl}}{\sqrt{12\pi}}\Rightarrow\phi(t)=\phi_0-\frac{mM_\text{Pl}}{\sqrt{12\pi}}\,t.$$ If substitute $\phi(t)$ from this last equation into the Friedmann equation in \eqref{feq-slr-app} we get that \cite{infl-linde-31}: 

\bea H=\pm\sqrt\frac{4\pi}{3}\frac{m}{M_\text{Pl}}\left(\phi_0-\frac{mM_\text{Pl}}{\sqrt{12\pi}}\,t\right)\Rightarrow a(t)=a_0\exp\left[\pm\sqrt\frac{4\pi}{3}\frac{m\phi_0}{M_\text{Pl}}\,t\left(1-\frac{mM_\text{Pl}}{2\sqrt{12\pi}\,\phi_0}\,t\right)\right].\label{a-exp}\eea For sufficiently small $t\ll 2\sqrt{12\pi}\phi_0/mM_\text{Pl}$, the above evolution law represents de Sitter expansion: $a(t)\approx a_0\exp(Ht)$ with $H=H_0\approx\sqrt{4\pi/3}\,m\phi_0/M_\text{Pl}$. Hence, since the intial value $\phi_0$ must obey the bonds \eqref{phi0-bonds}, the typical time at which the Universe considerably expands, $$\delta t\sim H_0^{-1}=\sqrt\frac{3}{4\pi}\frac{M_\text{Pl}}{m\phi_0}\;\Rightarrow\;\frac{1}{M_\text{Pl}}\lesssim\delta t\ll\frac{3}{m},$$ so that we get the following estimated bonds: 

\bea \frac{m^3}{3M^3_\text{Pl}}\lesssim|\delta z|\ll\frac{m^2}{M^2_\text{Pl}}\;\Rightarrow\;10^{-15}\lesssim|\delta z|\ll 10^{-10},\label{estimates}\eea i. e., $10^{-13}\,\%\lesssim RP\ll 10^{-8}\,\%$, are the bonds for the relative probability of initial conditions that lead to slow-roll inflation in this model. This result puts the minimally-coupled (single field) $\phi^2$-inflation in serious travels as a competitive model for primordial inflation and contradicts previous claims on the great generality of inflationary regimes in this model \cite{zeldovich-1985}.


\begin{figure}[t!]\begin{center}
\includegraphics[width=7cm]{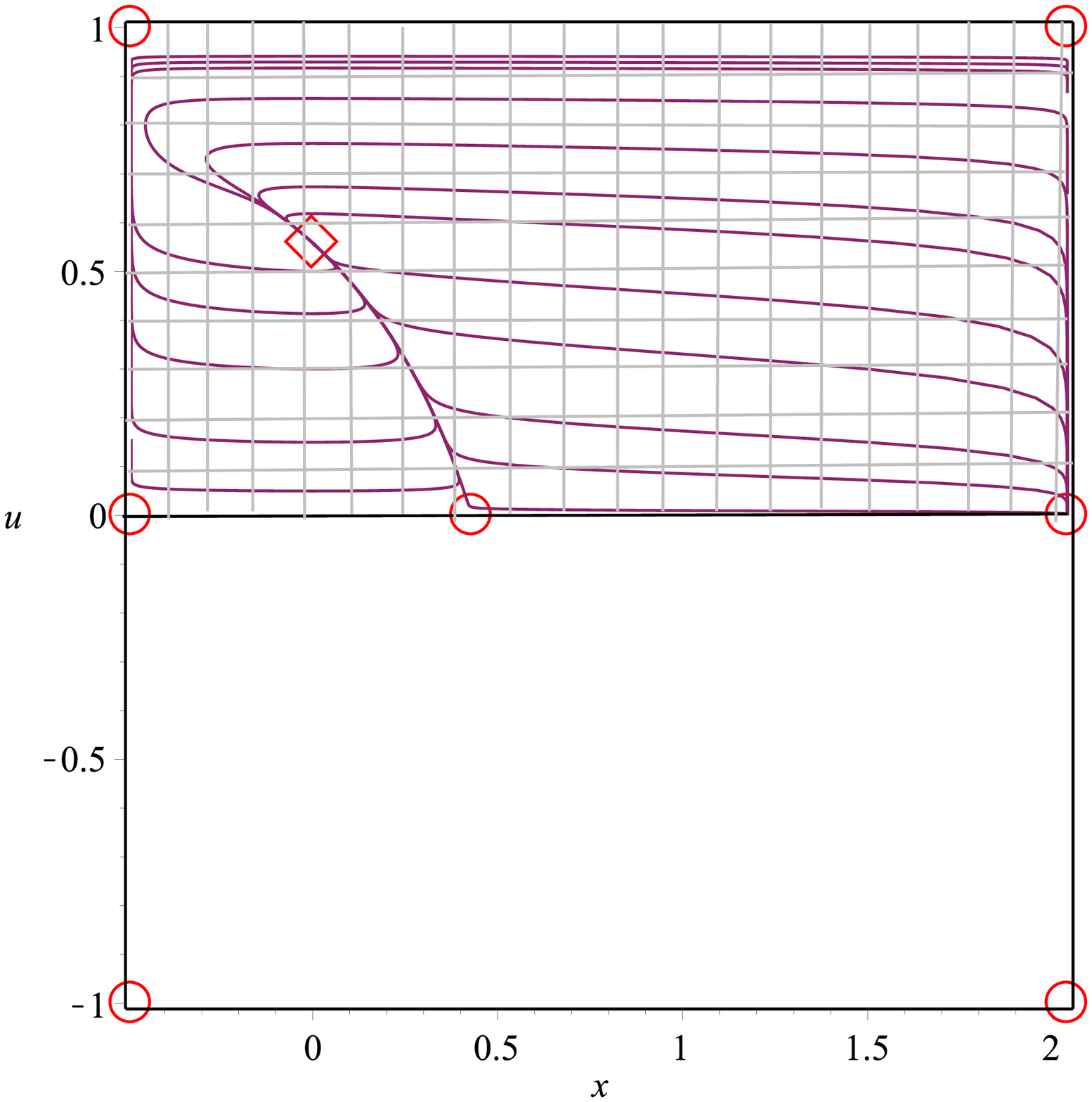}
\includegraphics[width=7cm]{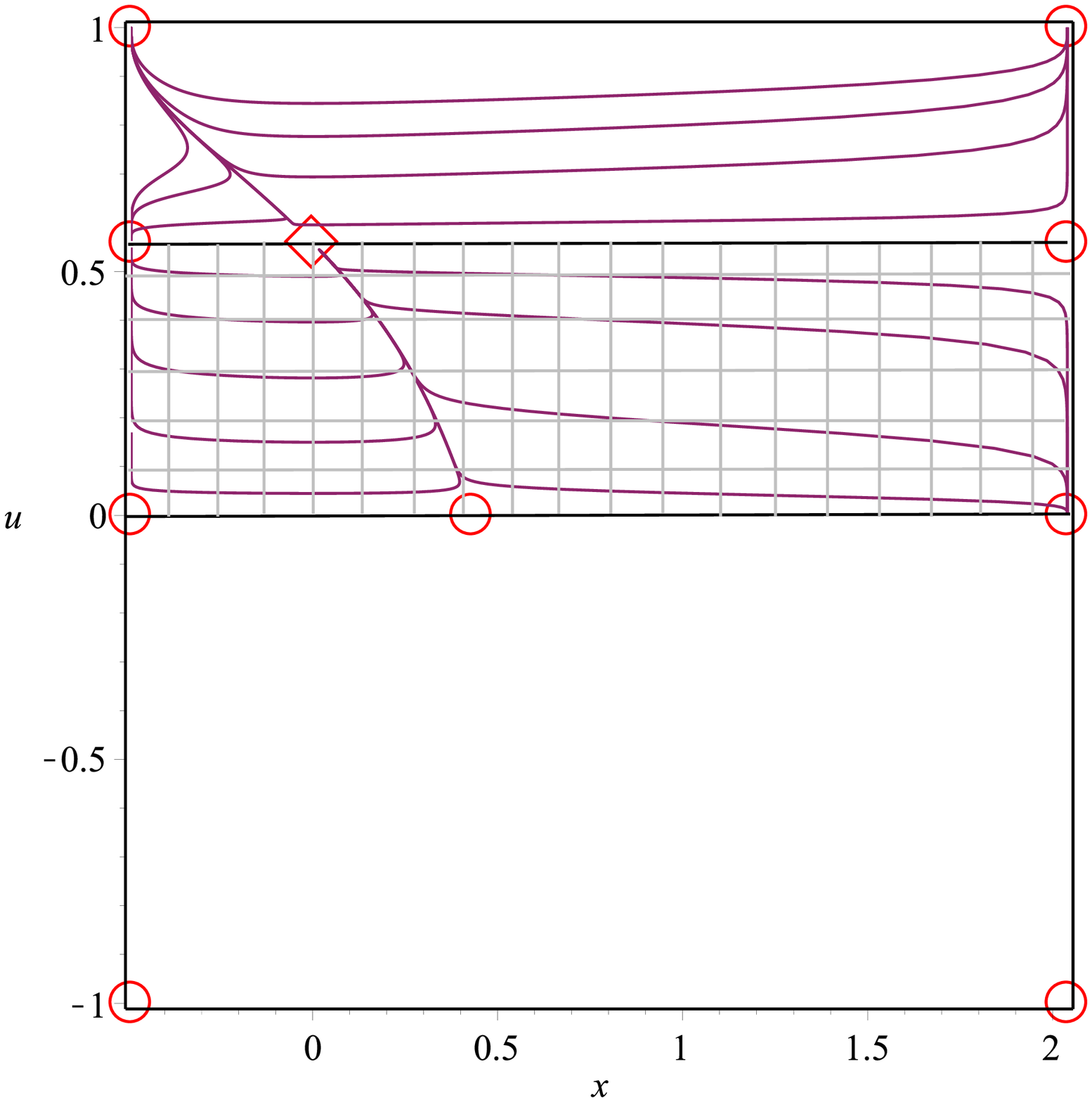}
\vspace{1.2cm}\caption{Towards estimating the relative amount of initial data leading to slow-roll inflation in the BD theory with exponential potential -- left hand panel -- and with symmetry-breaking potential -- right hand panel. Both panels coincide with the corresponding ones in FIG. \ref{fig3}, where the phase portraits were drawn for $\epsilon=0.1$. In the present drawings the curves corresponding to the slow-roll conditions as well as the phase space orbits in the lower half of the phase plane, have been conveniently removed. The gridded rectangular domains correspond to the regions of the phase space where the initial conditions lead to the slow-roll de Sitter local attractor (the diamond) in each case.}\label{fig6}\end{center}\end{figure}



Let us check now other models such as the BD theory with the exponential and with the symmetry-breaking type of potentials (see the phase portraits in FIG. \ref{fig3} and see also FIG. \ref{fig6} where the curves representing the slow-roll conditions, as well as the orbits in the lower half of the phase plane, have been conveniently removed). For these cases the phase plane is the rectangle:

\bea \Psi_\text{BD}=\left\{\left(x,u\right)|a_-\leq x\leq a_+,-1\leq u\leq 1\right\},\label{bd-rectang}\eea where $a_\pm=\sqrt{6\epsilon}\pm\sqrt{6\epsilon+1}$ and $u=v_+\cup v_-$ with $v_\pm$ defined in \eqref{z-var}. The area of this rectangle: $\text{Area}(\Psi_\text{BD})=\Delta u\times\Delta x=2\times\sqrt{6\epsilon+1}$, where $\Delta u=2$ is the height and $\Delta x=\sqrt{6\epsilon+1}$ is the width of the rectangle. Take first the BD model with the exponential potential. In this case, as seen from the left hand panel of FIG. \ref{fig3} (and of FIG. \ref{fig6}), the slow-roll de Sitter critical point is a local attractor in the upper half of the phase rectangle, while in the lower half it is the generic stiff-matter critical point $P^+_\text{g-stiff,-}:(a_+,-1)$ the local attractor. This means that any initial condition in the upper half of the rectangle inevitably leads to the de Sitter local attractor, while any initial condition given in the lower half leads to the stiff-matter solution. Hence, the area of the upper rectangle (gridded area in the left hand panel of FIG. \ref{fig6}) is given by $\text{Area}(\Psi^\uparrow_\text{BD})=1\times\sqrt{6\epsilon+1}$, so that the relative probability:

\bea RP=\frac{\text{Area}(\Psi^\uparrow_\text{BD})}{\text{Area}(\Psi_\text{BD})}\times 100\,\%=50\,\%.\label{exp-rp}\eea 

For the BD theory with the symmetry-breaking potential the situation is a bit more complex. In this case, as seen from FIG. \ref{fig3} and FIG. \ref{fig6} (right hand panel), the slow-rolling de Sitter critical point (enclosed by the diamond) is a local attractor only for orbits that lie in the region $\Psi^*_\text{BD}=\{(x,u)|a_-\leq x\leq a_+,0\leq u\leq 4\sqrt\epsilon/(4\sqrt\epsilon+1)\}$ (gridded strip in the right hand panel of FIG. \ref{fig6}). For orbits in the region above this strip the de Sitter solution is a saddle point. Then one has to apply the same approach followed in the minimally-coupled $\phi^2$-inflation model. One defines a $\rho$-ball around the saddle point and then one estimates its radius. The results do not differ too much from the ones in the minimally-coupled $\phi^2$-inflation case and the relative probability is almost vanishing, so that, in order to estimate the relative amount of initial data leading to slow-roll inflation for the symmetry-breaking potential, it is enough to consider that all of the successful initial conditions are in the strip $\Psi^*_\text{BD}$ (gridded rectangular region in the right hand panel of FIG. \ref{fig6}). The area of this strip $$\text{Area}(\Psi^*_\text{BD})=\frac{4\sqrt\epsilon}{4\sqrt\epsilon+1}\times\sqrt{6\epsilon+1},$$ where $4\sqrt\epsilon/(4\sqrt\epsilon+1)$ is the height of the strip, so that the relative probability of slow-roll inflation is given by:

\bea RP=\frac{\text{Area}(\Psi^*_\text{BD})}{\text{Area}(\Psi_\text{BD})}\times 100\,\%=\frac{2\sqrt\epsilon}{4\sqrt\epsilon+1}\times 100\,\%.\label{sbpot-rp}\eea Recall that $\text{Area}(\Psi_\text{BD})=2\times\sqrt{6\epsilon+1}$. Hence, the relative probability of inflation depends on the value of the coupling parameter. For the BD theory with a non-interacting scalar field, i. e., with vanishing self-interaction potential, the bounds from Solar system experiments yield that $\epsilon\lesssim 10^{-5}$, so that $RP\sim 1\,\%$. However, for a self-interacting BD scalar field, if take into account the Chameleon effect \cite{khouri-cham, brax-cham, mota-cham}, the coupling constant can be of order unity $\epsilon\approx 1$ \cite{quiros-prd-2015-cham}, which leads to $RP\sim 40\,\%$. 


For the quartic potential (FIG. \ref{fig4}), since the de Sitter solution is a global attractor, the relative probability of inflation is $RP=100\,\%$. Notice that in this case the slow-roll inflation critical point may not be associated with primordial inflation but rather with the present stage of the cosmic expansion of the Universe. Actually, in this case if add matter with energy density $\rho_m$ and pressure $p_m$, instead of a cosmological constant, before the slow-roll de Sitter attractor is approached, given phase space orbits evolve in the vicinity of the saddle scaling point $P^\pm_{\text{sc},*}$, where $\dot\phi^2/\rho_m=$const. Besides, even if add a cosmological constant, the end point of the cosmic evolution is always the slow-roll inflationary solution instead of the de Sitter evolution one might associate with the cosmological constant. 


The above analysis leads us to conclude that non-minimal coupling, in particular of the kind one founds in the BD theory, appreciably improves the relative probability of inflation in comparison with minimal coupling models such as the $\phi^2$-inflation, thus rendering inflation a natural outcome of the cosmological expansion in scalar-tensor theories of gravity.


\subsection{Non chaotic dynamics of Brans-Dicke cosmological models}

 There has been a debate on whether chaos arises in the phase space of scalar-tensor cosmological models \cite{calzetta-cqg-1993, bombelli-ijmp-1998, giacomini-prd-2001, faraoni-cqg-2006, gunzig-chaos-1, gunzig-chaos-2}. The possibility of chaos in the dynamical systems corresponding to scalar-tensor cosmological models has been established numerically \cite{calzetta-cqg-1993} and also semi-analytically through a perturbative approach \cite{bombelli-ijmp-1998}. In these papers the dynamical system was written in the form of a Hamiltonian system. In \cite{gunzig-chaos-1, gunzig-chaos-2} the occurrence of chaos has been challenged through a general dynamical system approach to classical self-consistent scalar field cosmology in the framework of spatially flat FRW spacetimes, for arbitrary potentials and arbitrary non-minimal coupling. In \cite{faraoni-cqg-2006} the non occurrence of chaos in scalar-field cosmologies with arbitrary couplings and potentials has been confirmed through using the conformal transformations approach. Here we shall to confirm this conclusion for the Brans-Dicke theory with arbitrary potentials for the flat FRW cosmologies.

It is a well-known fact that, according to the Poincar\`e-Bendixson theorem \cite{p-b-theor-book-2, p-b-theor-book-3}, strange attractors do not arise in 2D dynamical systems, so that the corresponding dynamics do not undergo chaos. This has been, precisely, the basis of the demonstration in \cite{gunzig-chaos-1, gunzig-chaos-2} of the absence of chaos in flat FRW scalar-tensor cosmological models. According to the present study this result is confirmed for vacuum Brans-Dicke theory. Actually, as shown in section \ref{sect-bd}, the BD equations of motion for vacuum (see \eqref{bd-mot-eq}):

\bea &&H^2=\frac{1}{6\epsilon}\left(\frac{\dot\phi}{\phi}\right)^2-2\frac{\dot\phi}{\phi}\,H+\frac{V}{3\epsilon\phi^2},\nonumber\\
&&\dot H=-\frac{1}{2\epsilon}\left(\frac{\dot\phi}{\phi}\right)^2+4H\frac{\dot\phi}{\phi}-\frac{4V-\phi V_\phi}{(1+6\epsilon)\phi^2},\nonumber\\
&&\frac{\ddot\phi}{\phi}+3H\frac{\dot\phi}{\phi}+\left(\frac{\dot\phi}{\phi}\right)^2=\frac{4V-\phi V_\phi}{(1+6\epsilon)\phi^2},\nonumber\eea may be traded by the 2D dymensional system \eqref{ode-ind-grav}:

\bea &&x'=\sqrt\frac{3}{2}\,(1+2\sqrt{6\epsilon}\,x-x^2)\left[-\sqrt{6}\,x+\frac{(1+\sqrt{6\epsilon}\,x)(4\sqrt\epsilon-z)}{1+6\epsilon}\right],\nonumber\\
&&z'=\sqrt{6}\,xz(\alpha z+\sqrt{\epsilon}),\nonumber\eea on the state space variables: $$x\equiv\frac{\dot\phi}{\sqrt{6\epsilon}\,\phi H},\;z\equiv\sqrt\epsilon\frac{\phi V_\phi}{V}.$$ In the above dynamical system $\alpha=\Gamma_V-1$ contains the information on the functional form of the potential. For a wide class of potentials $\alpha$ is either a number or a function of the variable $z$. For instance, for the exponential potential $\alpha=0$, while for the power-law potential $V\propto\phi^{2n}$, $\alpha=-1/n$. For the symmetry-breaking potential \eqref{s-b-pot}: $$\alpha=\alpha(z)=\frac{2\sqrt\epsilon-z}{2z},$$ etc. Hence, in those cases there is no chaos in the asymptotic dynamics of the flat FRW BD cosmological models.


\section{Conclusion}\label{sect-conclu}

In this paper we have done a quite general study of the phase space dynamics of single-field scalar-tensor cosmological models with arbitrary couplings and self-interaction potentials. In order to gain more insight we have specialized to NMC and BD theories. In the latter case a detailed study of the phase space dynamics has exposed the global dynamics of the vacuum for a wide variety of potentials and, also, the global dynamics of BD theory with a cosmological constant for the power-law type of potentials. Especial attention has been paid to the choice of the variables of the state space so that the global dynamics could be exposed. We have complemented the study with numeric calculations for specific potentials that made possible to draw the corresponding phase portraits.

What is more relevant: we have been able to give rough quantitative estimates of the relative probability of slow-roll inflation in several models, including the $\phi^2$-inflation model and Brans-Dicke models with the exponential and with the symmetry-breaking potential, respectively. Our results indicate that the non-minimal coupling of the scalar field to the curvature in the STT-s appreciably improves the bounds on the allowed initial data-set, pointing to the naturalness of slow-inflation in what regards to the allowed space of initial conditions. This is to be contrasted with the very strong bounds on minimal coupling theories, where the relative probability of slow-roll inflation is almost vanishing. This conclusion has been possible thanks to our choice of measure that enabled us to give quantitative estimates. In forthcoming work we want to perform a similar study for multi-field models of inflation.


\section*{Acknowledgments}

The authors thank SNI-CONACyT for continuous support of their research activity. UN acknowledges PRODEP-SEP and CIC-UMSNH for financial support of his contribution to the present research. RDA also acknowledges CONACyT for the postdoc grant 350411 under which part of this work was performed. The work of RGS was partially supported by SIP20200666, COFAA-IPN, and EDI-IPN grants.


\end{document}